\documentclass{article}\usepackage[]{graphicx}\usepackage[]{color}
\makeatletter
\def\maxwidth{ %
  \ifdim\Gin@nat@width>\linewidth
    \linewidth
  \else
    \Gin@nat@width
  \fi
}
\makeatother

\definecolor{fgcolor}{rgb}{0.345, 0.345, 0.345}
\newcommand{\hlnum}[1]{\textcolor[rgb]{0.686,0.059,0.569}{#1}}%
\newcommand{\hlstr}[1]{\textcolor[rgb]{0.192,0.494,0.8}{#1}}%
\newcommand{\hlcom}[1]{\textcolor[rgb]{0.678,0.584,0.686}{\textit{#1}}}%
\newcommand{\hlopt}[1]{\textcolor[rgb]{0,0,0}{#1}}%
\newcommand{\hlstd}[1]{\textcolor[rgb]{0.345,0.345,0.345}{#1}}%
\newcommand{\hlkwa}[1]{\textcolor[rgb]{0.161,0.373,0.58}{\textbf{#1}}}%
\newcommand{\hlkwb}[1]{\textcolor[rgb]{0.69,0.353,0.396}{#1}}%
\newcommand{\hlkwc}[1]{\textcolor[rgb]{0.333,0.667,0.333}{#1}}%
\newcommand{\hlkwd}[1]{\textcolor[rgb]{0.737,0.353,0.396}{\textbf{#1}}}%

\usepackage{framed}
\makeatletter
\newenvironment{kframe}{%
 \def\at@end@of@kframe{}%
 \ifinner\ifhmode%
  \def\at@end@of@kframe{\end{minipage}}%
  \begin{minipage}{\columnwidth}%
 \fi\fi%
 \def\FrameCommand##1{\hskip\@totalleftmargin \hskip-\fboxsep
 \colorbox{shadecolor}{##1}\hskip-\fboxsep
     \hskip-\linewidth \hskip-\@totalleftmargin \hskip\columnwidth}%
 \MakeFramed {\advance\hsize-\width
   \@totalleftmargin\z@ \linewidth\hsize
   \@setminipage}}%
 {\par\unskip\endMakeFramed%
 \at@end@of@kframe}
\makeatother

\definecolor{shadecolor}{rgb}{.97, .97, .97}
\definecolor{messagecolor}{rgb}{0, 0, 0}
\definecolor{warningcolor}{rgb}{1, 0, 1}
\definecolor{errorcolor}{rgb}{1, 0, 0}
\newenvironment{knitrout}{}{} 

\usepackage{alltt}
\usepackage{graphicx, fancyhdr, amssymb, amsmath, amsthm,epsfig, mathrsfs, verbatim, setspace,float,algorithmicx,algorithm,algpseudocode,bm,blkarray,mathabx,enumitem}
\usepackage{natbib}
\renewenvironment{knitrout}{\begin{singlespace}}{\end{singlespace}}
\usepackage[margin=0.75in]{geometry}

\newcommand{\utwi}[1]{\mbox{\boldmath $ #1$}}
\newcommand{\Y}{{\utwi{Y}}}
\newcommand{\X}{{\utwi{X}}}
\newcommand{\B}{{\utwi{B}}}
\newcommand{\Z}{{\utwi{Z}}}
\newcommand{\PhiB}{{\utwi{\Phi}}}

\newcommand{\BigVAR}{{\tt BigVAR }}
\newcommand{\betaB}{{\utwi{\beta}}}
\DeclareMathOperator{\vect}{vec}
\newcommand{\mLabel}[1]{\mbox{$\scriptstyle{#1}$}}

\newcommand{\argmin}{\operatornamewithlimits{argmin}}

\IfFileExists{upquote.sty}{\usepackage{upquote}}{}
\begin{document}

\title{BigVAR: Tools for Modeling Sparse High-Dimensional Multivariate Time Series\footnote{The development of BigVAR was supported by a 2014 Google Summer of Code scholarship.  This research was supported by an Amazon Web Services in Education Research Grant.  DSM was supported by a Xerox PARC Faculty Research Award and NSF Grant DMS-1455172.  JB was supported by NSF DMS-1405746.
}
}
\author{William B. Nicholson\footnote{
Corresponding Author,
PhD Candidate,  
Department of Statistical Science,
Cornell University,
301 Malott Hall,
Ithaca, NY 14853
(E-mail: \href{mailto:wbn8@cornell.edu}{wbn8@cornell.edu}; Webpage: \url{http://www.wbnicholson.com})},
  David S. Matteson\footnote{ 
Assistant Professor, 
Department of Statistical Science and Department of Social Statistics,
Cornell University,
1196 Comstock Hall,
Ithaca, NY 14853,
(E-mail: \href{mailto:matteson@cornell.edu}{matteson@cornell.edu}; Webpage: \url{https://courses.cit.cornell.edu/\textasciitilde dm484/})},
and Jacob Bien\footnote{
Assistant Professor, 
Department of Biological Statistics and Computational Biology and Department of Statistical Science,
Cornell University,
1178 Comstock Hall,
Ithaca, NY 14853
(E-mail: \href{mailto:jbien@cornell.edu}{jbien@cornell.edu}; Webpage: \url{http://faculty.bscb.cornell.edu/\textasciitilde bien/})}
}

\date{\today}
\maketitle
\begin{abstract}
The R package {\tt BigVAR} allows for the simultaneous estimation of high-dimensional time series by applying structured penalties to the conventional vector autoregression (VAR) and vector autoregression with exogenous variables (VARX) frameworks.  Our methods can be utilized in many forecasting applications that make use of time-dependent data such as macroeconomics, finance, and internet traffic.  Our package extends solution algorithms from the machine learning and signal processing literatures to a time dependent setting: selecting the regularization parameter by sequential cross validation and provides substantial improvements in forecasting performance over conventional methods.  We offer a user-friendly interface that utilizes {\tt R}'s s4 object class structure which makes our methodology easily accessible to practicioners. 

In this paper, we present an overview of our notation, the models that comprise {\tt BigVAR}, and the functionality of our package with a detailed example using publicly available macroeconomic data.  In addition, we present a simulation study comparing the performance of several procedures that refit the support selected by a {\tt BigVAR} procedure according to several variants of least squares and conclude that refitting generally degrades forecast performance.    
\end{abstract}
\section{Introduction}

For decades, the vector autoregression (VAR) and vector autoregression with unmodeled exogenous variables (VARX) have served as essential tools in forecasting multivariate time series.  However, in the absence of regularization, the VAR and VARX are heavily overparameterized, often forcing practitioners to arbitrarily specify a reduced subset of series to model. 

Recent years have witnessed tremendous developments toward the incorporation of regularization methods in the forecasting of high-dimensional multivariate time series with a particular interest in the lasso \citep{tibs} and its structured variants (the group lasso, \cite{yuan} and sparse group lasso, \cite{Simon}).  All of these methods can be expressed as penalized least squares optimization problems which can be solved efficiently with iterative nonsmooth convex optimization algorithms, such as coordinate descent \citep{Friedman} and generalized gradient descent \citep{beck}.

Despite growing interest in the area, there has been relatively little progress in the development of software that allows for the modeling of sparse high-dimensional VARs and VARXs.  Many authors, including \cite{Davis} and \cite{BickelSong} implement their penalized VAR models as modifications of the existing implementation {\tt glmnet} \citep{glmnet}, a package that is not designed for time-dependent problems and offers limited multivariate and structured support.   

Moreover, we have found a dearth of {\tt R} packages that even allow for the estimation of a high-dimensional VAR or VARX by least squares.  The {\tt ar.ols} function in base {\tt R} employs explicit matrix inversion, hence it is not tractable in high-dimensional settings and does not have VARX support.  The {\tt VAR} function in the package {\tt vars} fits the VAR equation-by-equation via least squares using {\tt lm}, which can cause complications under scenarios in which the number of covariates is close to or exceeds the length of the series, as such an implementation ignores degrees of freedom and can potentially can lead to numerically unstable results.

\BigVAR adapts the aforementioned penalized regression solution algorithms from the regularization literature to a multivariate time series setting, allowing for the simultaneous forecasting of many potentially interrelated time series.  
If forecasts are only desired from a subset of included series, \BigVAR utilizes the VARX-L framework \citep{NicholsonSR} to effectively leverage the information from unmodeled \emph{exogenous} series to improve the forecasts of modeled \emph{endogenous} series.

We additionally offer a class of Hierarchical Vector Autoregression (HVAR) procedures \citep{Nicholson} that address the notion of lag order by imposing a nested group lasso penalty in the VAR context.  Finally, for comparison purposes, we offer very fast and numerically stable implementations of information criterion based models which fit VAR and VARX models by least squares as the minimizer of either AIC or BIC.

Section \ref{sec2} details our notation and provides an overview of the VARX-L and HVAR frameworks and Section \ref{sec5} details the practical implementation of \BigVAR with a macroeconomic data example.  Section \ref{sec4} provides an overview of several post-estimation refitting procedures as well as a simulation study, and Section \ref{secconc} contains our conclusion.  Our appendix elaborates upon our solution methods and algorithms.

\section{Notation and Overview of \BigVAR Procedures}
\label{sec2}
 
Let $\{ \mathbf{y_t}\}_{t = 1}^T$ denote a $k$ dimensional vector time series and $\{\mathbf{x}_t\}_{t=1}^{T}$ denote an $m$-dimensional unmodeled \emph{exogenous} series.  A vector autoregression with exogenous variables of order (p,s) , VARX$_{k,m}$($p,s$), can be expressed as  
\begin{align}
  \label{VAR1}
\mathbf{y}_t=\utwi{\nu}+\sum_{\ell=1}^p\PhiB^{(\ell)}\mathbf{y}_{t-\ell}+\sum_{j=1}^s \betaB^{(j)}\mathbf{x}_{t-j}+\mathbf{u}_t \; \text{ for } \;t=1,\ldots,T,
\end{align}
in which $\utwi{\nu}$ denotes a $k\times 1$ intercept vector, each $\PhiB^{(\ell)}$ represents a $k\times k$ endogenous (modeled) coefficient matrix, each $\betaB^{(j)}$ represents a $k\times m$ exogenous (unmodeled) coefficient matrix, and $\mathbf{u}_t\stackrel{\text{wn}}{\sim}(\mathbf{0},\mathbf{\Sigma}_u)$.  Note the the VAR is a special case of Equation \eqref{VAR1} in which the second summation ($\sum_{j=1}^s \betaB^{(j)}\mathbf{x}_{t-j}$) is not included.

\subsection{The VARX-L Framework}

To reduce the parameter space of the VARX, the VARX-L framework applies structured convex penalties to the least squares VARX problem, resulting in the objective  
\begin{align}
\label{PenFunForm}
\min_{\utwi{\nu},\PhiB,\betaB}  \sum_{t=1}^T\|\mathbf{y}_t-\utwi{\nu}-\sum_{\ell=1}^p\PhiB^{(\ell)}\mathbf{y}_{t-\ell}-\sum_{j=1}^s \betaB^{(j)}\mathbf{x}_{t-j}\|_F^2+\lambda \bigg(\mathcal{P}_y(\PhiB)+ \mathcal{P}_x(\betaB)\bigg),
\end{align}
in which $\|A\|_F$ denotes the Frobenius norm of matrix A (i.e. the elementwise 2-norm), $\PhiB=[\PhiB^{(1)},\dots,\PhiB^{(p)}]$, $\betaB=[\betaB^{(1)},\dots,\betaB^{(s)}]$, $\lambda\geq 0$ is a penalty parameter estimated by sequential cross-validation, $\mathcal{P}_y(\PhiB)$ represents the group penalty structure on endogenous coefficients, and $\mathcal{P}_x(\betaB)$ represents the group penalty structure on exogenous coefficients.

These penalties impose structured sparsity based upon a partition of the parameter space that takes into account the intrinsic structure of the VARX.  All VARX-L penalty structures are detailed in Table \ref{tab:tabGS}.  Observe that groups are weighted by their cardinality to prevent regularization favoring larger groups.  Plots of example sparsity patterns (with nonzero, or \emph{active} coefficients shaded) are depicted in Figure \ref{fig:plots}.  In the following sections, we will describe each penalty structure in more detail.

\begin{knitrout}
\definecolor{shadecolor}{rgb}{0.969, 0.969, 0.969}\color{fgcolor}\begin{figure}[!h]
\includegraphics[width=\maxwidth]{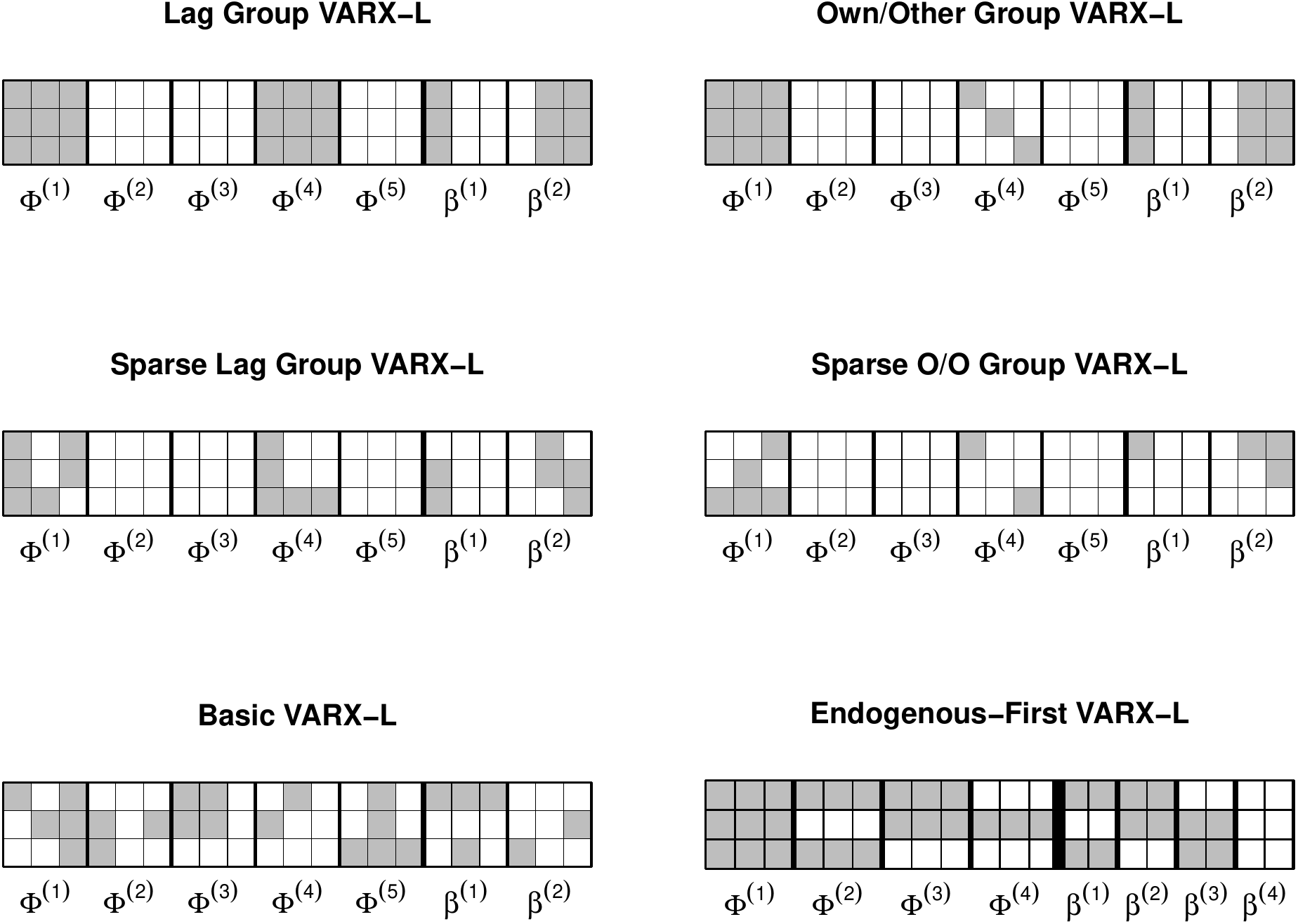} \caption[Examples of VARX-L Sparsity Patterns (k=3, p=5]{Examples of VARX-L Sparsity Patterns (k=3, p=5; m=2, s=3).  The gray shading denotes nonzero 'active' coefficients whereas white denotes coefficients that have been set to zero.}\label{fig:plots}
\end{figure}

\end{knitrout}

      \begin{table}[H]
     \centering    
\scriptsize
    \caption{\footnotesize \label{tab:tabGS} VARX-L Penalty Functions (Reproduced from \cite{NicholsonSR}).  Note that $\PhiB_{\text{on}}^{(\ell)}$ and $\PhiB_{\text{off}}^{(\ell)}$ denote the diagonal and off-diagonal elements of coefficient matrix $\PhiB^{(\ell)}$, respectively.  
    }  
   \begin{tabular}{ l | c| c }
Group Name& $\mathcal{P}_y(\PhiB)$ & $\mathcal{P}_x(\betaB)$ \\
    \hline
\refstepcounter{equation}(\theequation) \label{VARX1} Lag &  $\sqrt{k^2}\sum_{\ell=1}^p\|\PhiB^{(\ell)}\|_F$ &  $\sqrt{k}\sum_{j=1}^s\sum_{i=1}^m\|\betaB_{\cdot,i}^{(j)}\|_F$  \\
\refstepcounter{equation}(\theequation) \label{VARX2} Own/Other& $\sqrt{k}\sum_{\ell=1}^p ||\PhiB_{\text{on}}^{(\ell)}||_F+\sqrt{k(k-1)}\sum_{\ell=1}^p||\PhiB_{\text{off}}^{(\ell)}||_F$ &$\sqrt{k}\sum_{j=1}^s\sum_{i=1}^m\|\betaB_{\cdot,i}^{(j)}\|_F$ \\
\refstepcounter{equation}(\theequation) \label{VARX3} Sparse Lag& $(1-\alpha)\sqrt{k^2}\sum_{\ell=1}^p\|\PhiB^{(\ell)}\|_F+\alpha\|\PhiB\|_1$ & $(1-\alpha)\sqrt{k}\sum_{j=1}^s\sum_{i=1}^m\|\betaB_{\cdot,i}^{(j)}\|_F$ +$\alpha\|\betaB\|_1$  \\
\refstepcounter{equation}(\theequation) \label{VARX4} Sparse Own/Other& $(1-\alpha)\big(\sqrt{k}\sum_{\ell=1}^p||\PhiB_{\text{on}}^{(\ell)}||_F+\sqrt{k(k-1)}\sum_{\ell=1}^p||\PhiB_{\text{off}}^{(\ell)}||_F\big)+\alpha\|\PhiB\|_1$ &  $(1-\alpha)\sqrt{k}\sum_{j=1}^s\sum_{i=1}^m\|\betaB_{\cdot,i}^{(j)}\|_F$ +$\alpha\|\betaB\|_1$\\
\refstepcounter{equation}(\theequation) \label{VARX5} Basic & $\|\PhiB\|_1$  & $\|\betaB\|_1$ \\  
\hline\\
\refstepcounter{equation}(\theequation) \label{VARX6} Endogenous-First & \multicolumn{2}{c}{$\mathcal{P}_{y,x}(\PhiB,\betaB)=\sum_{\ell=1}^p\sum_{j=1}^k\bigg(\|[\PhiB_{j,\cdot}^{(\ell)},\betaB_{j,\cdot}^{(\ell)}]\|_F+\|\betaB_{j,\cdot}^{(\ell)}\|_F\bigg)$}   


  \end{tabular}
 \end{table}

\subsubsection*{Group Lasso Penalties}
The group lasso \citep{yuan} has emerged as a popular penalized regression procedure that partitions all model coefficients into a collection of disjoint groups that can take into account the inherent structure of a multivariate time series.  Within a group, all coefficients will either be set to zero or the group will be \emph{active} and all coefficients will be nonzero.  We consider two group structures for the endogenous covariates: a lag based grouping (\emph{Lag} Group VARX-L, expression \ref{VARX1} in Table \ref{tab:tabGS}) and a grouping that distinguishes between a series' \emph{own} lags (diagonal entries of $\PhiB^{(\ell)}$) and those of other series (off diagonal entries of $\PhiB^{(\ell)}$) (\emph{Own/Other} Group VARX-L, expression \ref{VARX2}).  The Own/Other grouping incorporates the widely held stylized fact in macroeconometrics that a series' own lags have different dynamic dependence than those from other series \citep{Litterman1979}.  

Though both penalties employ the same solution algorithm, since the partitioning under the Lag Group VARX-L forms proper submatrices, it is possible to directly solve the matrix optimization problem as opposed to performing a least squares transformation, resulting in substantially less computational overhead than the Own/Other scenario.    

Both the Own/Other and Lag Group VARX-L partition exogenous coefficients by column.  Our experiences have found that assigning each exogenous covariate to its own group substantially increases computation time without an improvement in forecast performance and an exogenous lag-based grouping is too general.  Hence, the column-based grouping serves as a compromise; allowing for a degree of flexibility while still resulting in a computationally efficient optimization problem.    

\subsubsection*{Sparse Group Lasso Penalties}
In certain scenarios, a group penalty can be too restrictive.  If a group is active, all of its coefficients are potentially nonzero.  On the other hand, specifying a large number of groups will substantially increase computation time and, in our experience, generally does not lead to improvements in forecasting performance.   

As a compromise, we consider applying \emph{sparse group lasso} penalties (expressions \ref{VARX3} and \ref{VARX4} in Table \ref{tab:tabGS}) proposed by \cite{Simon}, which allow for ``within-group'' sparsity via a convex combination of $L_1$(unstructured sparsity) and $L_2$ (structured sparsity) penalties.  \BigVAR offers the Sparse VARX-L for both the Lag and Own/Other structured groupings.

By default $\alpha$, the parameter that sets the weights of the two penalties and is constrained to be between 0 and 1, is chosen to according to a heuristic ($\frac{1}{k+1}$) to control within-group sparsity.  \BigVAR also permits for the joint cross validation of $\lambda$ and $\alpha$.  Performing joint cross validation allows for the Sparse Group VARX-L to function as a powerful diagnostic tool to determine the applicability of a structured grouping.  A selected value of close to zero provides strong evidence of structured sparsity whereas a value close to one points to a lack of structure.  

\subsubsection*{Basic Penalty}
The Basic VARX-L (expression \ref{VARX5} in Table \ref{tab:tabGS}) is the most general grouping and can be viewed as partitioning each variable into its own group or as applying an unstructured lasso penalty to the entire VARX coefficient matrix.  It does not incorporate any of the structure of the VARX, but it results in a comparably simpler optimization problem, allowing it to scale to much larger problems than structured penalties.

\subsubsection*{Nested Penalty Structures}
The previous penalty structures are disjoint groupings that partition $[\PhiB,\betaB]$.  In certain scenarios, one might wish to assign a preference to endogenous versus exogenous variables.  The \emph{Endogenous-First} VARX-L (expression \ref{VARX6} in Table \ref{tab:tabGS}) utilizes a nested penalty to prioritize endogenous series.  At a given lag, an exogenous series can enter the model only if their endogenous counterpart is nonzero.  Note that by construction this penalty decouples across series, allowing for endogenous/exogenous dependence to vary.  It is additionally required that $p\geq s$, otherwise such a nested penalty structure would not be appropriate.   

\subsubsection*{Solution Methods}
In order to solve the optimization problems in the form of Equation \ref{PenFunForm}, we employ computationally tractable algorithms designed for non-smooth convex functions.  Our solution methods do not make calls to external packages or commercial convex solvers and are optimized for time dependent problems.  All of our solution algorithms are coded in {\tt C++} and linked to {\tt R} via {\tt Rcpp} \citep{Rcpp}, {\tt RcppArmadillo} \citep{Rcpparmadillo}, and {\tt RcppEigen} \citep{Rcppeigen}.  The specific algorithms that we utilize for each procedure are displayed in Table \ref{tab:tabSA} in Section \ref{tabalg} of the appendix.  Implementation details are provided in the appendix of \cite{NicholsonSR}.
 
\subsection{Hierarchical Vector Autoregression (HVAR)}
\label{sec3}

The VARX-L procedures remain agnostic with regard to lag order selection.  Hence, as the maximum lag order increases forecast performance may start to degrade, as each group is treated democratically despite more distant lags generally tending to be less useful in forecasting.  Within the VAR context, we utilize the HVAR class of models \citep{Nicholson} which alleviate this issue by embedding lag order into \emph{hierarchical group lasso} penalties.

In addition to returning sparse solutions, our $\text{HVAR}_k(p)$ procedures induce regularization toward models with low maximum lag order.  To allow for greater flexibility, instead of imposing a single, universal lag order (as information criterion minimization based approaches tend to do), we allow it to vary across marginal models (i.e. the rows of the coefficient matrix $\PhiB=[\PhiB^{(1)},\dots,\PhiB^{(p)}]$).  \BigVAR includes three HVAR models as well as the ``Lag-weighted Lasso,'' which incorporates a lasso penalty that increases geometrically as the lag order increases.  These procedures are presented in Table \ref{tab:tabGS2} and example sparsity patterns of the HVAR procedures and the Lag-weighted Lasso are depicted in Figure \ref{fig:HVARplots}.

\begin{knitrout}
\definecolor{shadecolor}{rgb}{0.969, 0.969, 0.969}\color{fgcolor}\begin{figure}[!h]
\includegraphics[width=\maxwidth]{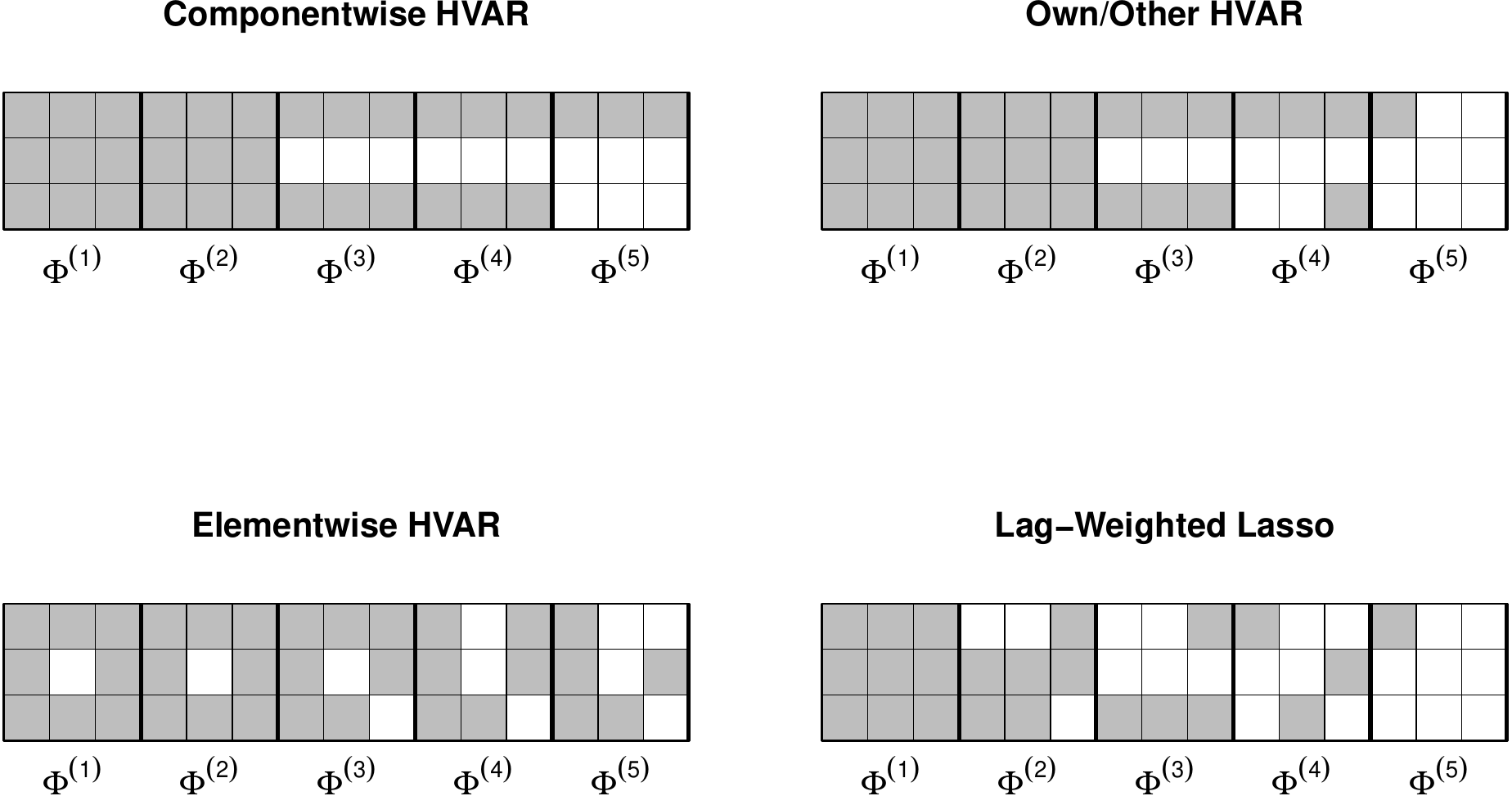} \caption[Examples of Sparsity Patterns for the HVAR procedures and the Lag-Weighted Lasso (k=3,p=5)]{Examples of Sparsity Patterns for the HVAR procedures and the Lag-Weighted Lasso (k=3,p=5)}\label{fig:HVARplots}
\end{figure}

\end{knitrout}

      \begin{table}[H]
     \centering    
    \caption{\footnotesize \label{tab:tabGS2} HVAR Penalty Functions . 
    }  
   \begin{tabular}{ l | c }
Group Name& $\mathcal{P}_y(\PhiB)$  \\
    \hline
\refstepcounter{equation}(\theequation) \label{HVAR1} Componentwise &  $\sum_{i=1}^k\sum_{\ell=1}^p\|\PhiB_i^{(\ell:p)}\|_2.$  \\
\refstepcounter{equation}(\theequation) \label{HVAR2} Own/Other& $\sum_{i=1}^k\sum_{\ell=1}^p\left[\|\PhiB_{i}^{(\ell:p)}\|_2+\|(\PhiB_{i,-i}^{(\ell)}, \PhiB_{i}^{([\ell+1]:p)})\|_2\right]$ \\
\refstepcounter{equation}(\theequation) \label{HVAR3} Elementwise& $\sum_{i=1}^k\sum_{j=1}^k\sum_{\ell=1}^p\|\PhiB_{ij}^{(\ell:p)}\|_2$  \\
\hline
\refstepcounter{equation}(\theequation) \label{HVAR4} Lag-weighted Lasso& $\sum_{\ell=1}^p\ell^{\gamma}\|\PhiB^{(\ell)}\|_1$  
  \end{tabular}
 \end{table}

\subsubsection{Componentwise HVAR}
The Componentwise HVAR (defined in expression \eqref{HVAR1} in Table \ref{tab:tabGS2}), allows for the maximum lag order to vary across marginal models, but within a series all components have the same maximum lag.  This structure allows for $k$ potentially different lag orders,  

\subsubsection{Own/Other HVAR}
The Own/Other HVAR (defined in expression \eqref{HVAR2} in Table \ref{tab:tabGS2}), is similar to the Componentwise HVAR, but imposes an additional layer of hierarchy within a lag: prioritizing coefficients of lagged values of the series of forecasting interest (i.e. ``own'' lags) over those of other series.  This penalty incorporates a common specification in the Bayesian VAR with a Minnesota Prior \citep{Litterman1979} that ``own'' lags are more informative for forecasting purposes than ``other'' lags,  
\subsubsection{Elementwise HVAR}
The Elementwise HVAR (defined in expression \eqref{HVAR3} in Table \ref{tab:tabGS2}) is the most general structure; in each marginal model, each series may have its own maximum lag.  Under this framework, there are $k^2$ possible lag orders

\subsubsection{Lag-weighted Lasso}
In addition, for comparison purposes we provide a \emph{Lag-weighted Lasso} (expression \eqref{HVAR4} in Table \ref{tab:tabGS2}), which consists of a lasso penalty that increases geometrically with lag; $\gamma\in [0,1]$ is an additional penalty parameter that is jointly estimated with $\lambda$ according to sequential cross validation.  This is similar to the approach proposed by \cite{BickelSong}.  Though it encourages greater regularization at more distant lags, it does not explicitly force sparsity and requires the specification of an arbitrary functional form as well as an additional penalty parameter.

 \subsection{Penalty Parameter Selection}
 In order to account for time dependence, selection of the penalty parameter $\lambda$ is conducted in a rolling manner.  The penalty parameter, $\hat{\lambda}$, is selected from a grid of values $\lambda_1,\dots,\lambda_n$.  We perform sequential cross validation between times $T_1-h+1$ and $T_2-h+1$, in which $h$ denotes forecast horizon.     At $T_1-h+1$, we forecast $\hat{\mathbf{y}}_{T_1+h}^{\lambda_i}$ for $i=1,\dots,n$, and sequentially add observations until time $T_2-h+1$.  $T_2-h+2$ through $T-h+1$ is used for out of sample forecast evaluation.

Unless otherwise specified, \BigVAR sets $T_1=\left \lfloor \frac{T}{3} \right\rfloor,  T_2=\left\lfloor \frac{2T}{3} \right\rfloor$.  We choose $\hat{\lambda}$ as the minimizer of h-step ahead MSFE:

\begin{align*}
  MSFE(\lambda_i)=\frac{1}{(T_2-T_1-h+1)}\sum_{t=T_1-h+1}^{T_2-h} \|\hat{\mathbf{y}}_{t+h|t}^{\lambda_i}-\mathbf{y}_{t+h}\|_2^2,
\end{align*}

In the VAR context, there are two possible methods to obtain multi-step ahead forecasts: iterated one-step ahead predictions or directly forecasting the longer horizon.  Per \cite{clark}, in the VAR context, iterated h-step ahead forecasts have the form:
\begin{align*}
  \hat{\mathbf{y}}_{t+h|t}=\hat{\nu}+\sum_{\ell=1}^p\widehat{\PhiB}^{(\ell)}\hat{\mathbf{y}}_{t+h-i|t},
\end{align*}
whereas the alternative involves directly forecasting h-step ahead forecasts
\begin{align*}
  \hat{\mathbf{y}}_{t+h}=\hat{\nu}+\sum_{\ell=1}^p\widehat{\PhiB}^{(\ell)}\mathbf{y}_{t+1-i}.
\end{align*}
Both approaches have advantages; as noted by \cite{swforecast}, the direct approach could provide more accurate forecasts if the VAR is misspecified, however, if the model is correctly specified, the iterated approach is theoretically more efficient.  In the VAR setting, \BigVAR allows for the choice of either iterated or direct forecasts when optimizing over forecasts horizons greater than one.  In the VARX setting only direct forecasts are available, since we do not return forecasts of exogenous series.   

If the user wishes to employ their own penalty parameter selection routine, they can do so by calling {\tt BigVAR.est} within their code.  This procedure will be discussed in Section \ref{BVEST}.

\section{Forecasting VAR(X) models with {\tt BigVAR} }
\label{sec5}

In this section, we demonstrate how to utilize \BigVAR to forecast a set of quarterly macroeconomic indicators procured from the St. Louis Federal Reserve Economic Database (FRED) via Quandl.  We consider forecasting four US macroeconomic series: 
\begin{enumerate}[label=(\roman*)]
\item Consumer Price Index (CPI),
\item Federal Funds Rate (FFR),
\item Gross Domestic Product (GDP),
\item M1 (a measure of the liquid components of the money supply).  
\end{enumerate}

We first download the data using the API provided in the {\tt Quandl} package and then transform each series to stationarity by taking the log difference of CPI, M1, and GDP and the log of FFR (since it is already expressed as a rate).  The {\tt R} code that reproduces this analysis is available at \url{http://www.wbnicholson.com/BigVARDemo.R}.

The GDP and CPI series start in Quarter 1 of 1947, but since the Federal Funds Rate was not officially published until 1954 and M1 was not recorded until 1959, we discard all realizations of GDP and CPI before Quarter 3 of 1959.  The data ranges through Quarter 2 of 2015, resulting in $T=224$.  As is standard in the regularization framework, before estimation we standardize each series to have zero mean and unit variance.        
\begin{knitrout}
\definecolor{shadecolor}{rgb}{0.969, 0.969, 0.969}\color{fgcolor}\begin{figure}
\includegraphics[width=\maxwidth]{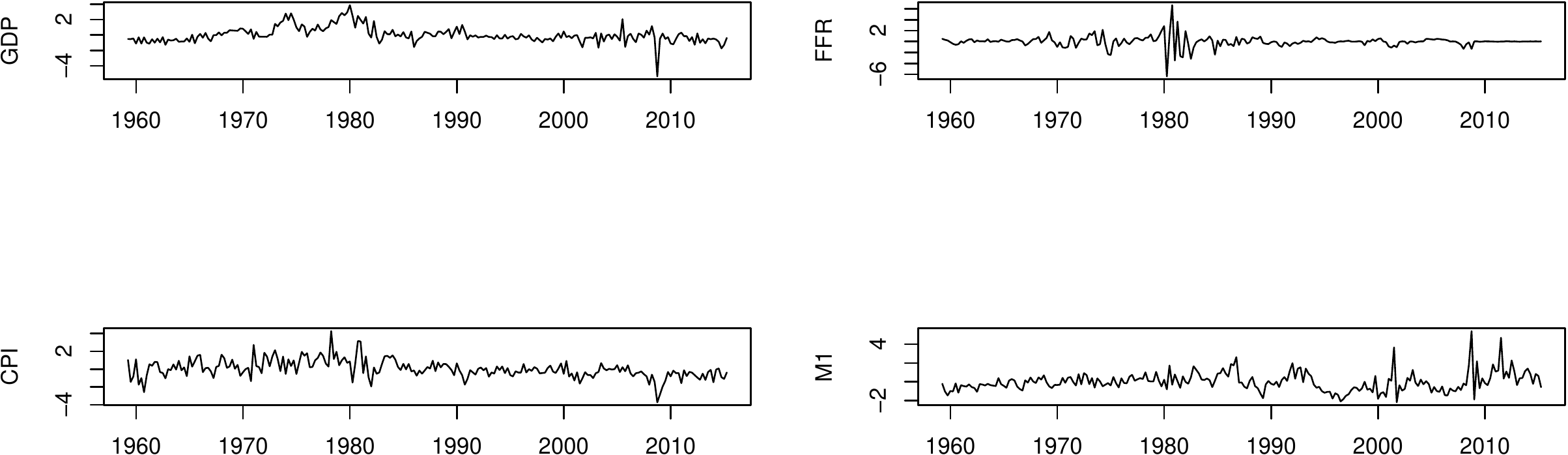} \caption[Plots of Standardized Quarterly GDP, Federal Funds Rate, CPI, and M1]{Plots of Standardized Quarterly GDP, Federal Funds Rate, CPI, and M1}\label{fig:macroplots}
\end{figure}

\end{knitrout}

\subsection{Constructing an object of class {\tt BigVAR}}
In an effort to streamline functionality, {\tt BigVAR} incorporates {\tt R}'s $s4$ object class system \cite{chamberss4}.  In order to fit a model, the user constructs an object of class {\tt BigVAR} that contains the data as well as model specifications.  A {\tt BigVAR} object can be created with the wrapper function {\tt constructModel}, which encompasses both the HVAR and VARX-L frameworks.

In examining Figure \ref{fig:macroplots}, we observe considerable fluctuations in the CPI and FFR series in the early 1980s, owing to the period's rapid inflation and the resulting contractionary monetary policy.  Here, we choose to minimize the influence of this period when selecting our penalty parameters.  Below, we construct an Elementwise $\text{HVAR}_{4}(4)$ and use data from Quarter 1 of 1985 to Quarter 1 of 2005 for penalty parameter selection.


\begin{knitrout}
\definecolor{shadecolor}{rgb}{0.969, 0.969, 0.969}\color{fgcolor}\begin{kframe}
\begin{alltt}
\hlkwd{library}\hlstd{(BigVAR)}

\hlstd{T1} \hlkwb{<-} \hlkwd{which}\hlstd{(}\hlkwd{index}\hlstd{(Y)}\hlopt{==}\hlstr{"1985 Q1"}\hlstd{)}
\hlstd{T2} \hlkwb{<-} \hlkwd{which}\hlstd{(}\hlkwd{index}\hlstd{(Y)}\hlopt{==}\hlstr{"2005 Q1"}\hlstd{)}

\hlstd{Model1}\hlkwb{=}\hlkwd{constructModel}\hlstd{(}\hlkwd{as.matrix}\hlstd{(Y),}\hlkwc{p}\hlstd{=}\hlnum{4}\hlstd{,}
    \hlkwc{struct}\hlstd{=}\hlstr{"HVARELEM"}\hlstd{,}\hlkwc{gran}\hlstd{=}\hlkwd{c}\hlstd{(}\hlnum{25}\hlstd{,}\hlnum{10}\hlstd{),}\hlkwc{verbose}\hlstd{=}\hlnum{FALSE}\hlstd{,}\hlkwc{VARX}\hlstd{=}\hlkwd{list}\hlstd{(),}\hlkwc{T1}\hlstd{=T1,}\hlkwc{T2}\hlstd{=T2)}
\end{alltt}
\end{kframe}
\end{knitrout}

The required arguments for {\tt constructModel} are: 
\begin{itemize}
\item {\tt Y}: a $T\times k$ multivariate time series (in matrix form),
\item {\tt p}: predetermined maximum lag order, 
\item {\tt gran}: two arguments that characterize the grid of penalty parameters: the first denotes the depth of the grid and the second the number of candidate penalty parameters.  
\end{itemize}
  
The choices for the argument {\tt struct} are presented in Table \ref{tab:tabStruct} in Section \ref{tabalg} in the appendix.  In the {\tt BigVAR} framework, {\tt gran} denotes the only ``hyperparameter'' that must be set by the end user.  Following \cite{Friedman}, the grid of penalty values starts with the smallest value in which all coefficients will be zero, then decrements in log linear increments.  The grid ends at a fraction of this maximum value (as dictated by the first argument in {\tt gran}).  These bounds are detailed in the appendix of \cite{NicholsonSR}.

In practice these bounds can be coarse. Consequently, to avoid scenarios in which several candidate penalty parameters return coefficient matrices of identically zero, \BigVAR utilizes an empirical procedure to determine tighter bounds.  In order to do so, we expand upon the approach presented in Algorithm 3 of \cite{bien1}.  Starting with the theoretically determined bound, we employ a bisection routine in order to find a tighter data-driven bound.  Our implementation of this procedure is detailed in Algorithm \ref{alg1} in Section \ref{tabalg} in the appendix.  In practice, we find that the best choices for grid depth tend to be between 10 and 50, depending on the number of series included and the forecast horizon.

The number of penalty parameters is also left to user input.  The package {\tt glmnet} calls for 100 penalty parameters by default.  However, in our applications we have found no substantial forecasting improvement in considering any more than 10.  If the user wishes to provide their own penalty parameters, they can do so through {\tt gran}, but they must also set the optional argument {\tt ownlambdas} to {\tt TRUE}.  The additional optional arguments to {\tt constructModel} and their default values are: 
\begin{itemize}
\item   {\tt RVAR}: Relaxed VAR(X) indicator to refit based upon the coefficients recovered from a VARX-L or HVAR procedure according to least squares (default: {\tt FALSE}).  This method will be discussed in greater detail in Section \ref{sec4}.
\item  {\tt MN}: option for the \emph{Minnesota VAR(X)}, which shrinks parameter estimates toward a vector random walk (default {\tt FALSE}).
\item {\tt h}: forecast horizon (default $1$).
\item {\tt verbose}: indicator for progress bar (default {\tt TRUE}).
\item {\tt IC}: indicator to return AIC and BIC benchmarks (default {\tt TRUE}).
\item {\tt VARX}: list of VARX specifications (default {\tt list()}).
\item {\tt T1}: start of cross validation period (default $\lfloor \frac{T}{3}\rfloor$).
\item {\tt T2}: start of forecast evaluation period (default $\lfloor \frac{2T}{3}\rfloor$).
\item {\tt ONESE}: indicator for \emph{One Standard Error} heuristic described in \cite{esl} which selects the largest penalty parameter within one standard error of the minimizer of MSFE (default {\tt FALSE}).
\item {\tt recursive}: indicator determining if recursive multi-step predictions are desired as opposed to direct (default {\tt FALSE}, applicable only for VAR models with $h>1$).  
\item {\tt alpha}: vector of candidate values for $\alpha$ if dual cross validation is desired for the Sparse Lag or Sparse Own/Other structured penalties (all entries must be between 0 and 1, the default value is $\frac{1}{k+1}$).
\item {\tt C}: vector denoting series to be shrunk toward a random walk instead of toward zero (used in situations in which some series exhibit signs of nonstationarity, while others don't).  This scenario will be discussed in greater detail in Section \ref{secMN}, (default $\mathbf{0}_k$, applicable only if {\tt MN} is {\tt TRUE}).
  
\end{itemize}  
  

\subsection{Implementation}
\label{IMP1}
In order to fit a model with \BigVAR using rolling cross validation, we simply need to execute the method {\tt cv.BigVAR} on an object of class \BigVAR as detailed below.  To fit an Elementwise $\text{HVAR}_{4}(4)$, we simply run the command 
\begin{knitrout}
\definecolor{shadecolor}{rgb}{0.969, 0.969, 0.969}\color{fgcolor}\begin{kframe}
\begin{alltt}
\hlstd{Model1Results} \hlkwb{=} \hlkwd{cv.BigVAR}\hlstd{(Model1)}
\end{alltt}
\end{kframe}
\end{knitrout}

An object of class {\tt BigVAR.Results} is returned.  By default, the output displays model characteristics, such as the penalty structure, maximum lag order, the value of $\lambda$ selected by rolling cross validation, and both in-sample and out-of-sample MSFE.  For comparison purposes, the out-of-sample MSFE from several benchmarks, including the sample mean, random walk, and the least squares VAR or VARX with lags selected by AIC and BIC are also returned.  

\begin{knitrout}
\definecolor{shadecolor}{rgb}{0.969, 0.969, 0.969}\color{fgcolor}\begin{kframe}
\begin{alltt}
\hlstd{Model1Results}
\end{alltt}
\begin{verbatim}
## *** BIGVAR MODEL Results *** 
## Structure
## [1] "HVARELEM"
## Forecast Horizon 
## [1] 1
## Minnesota VAR
## [1] FALSE
## Maximum Lag Order 
## [1] 4
## Optimal Lambda 
## [1] 6.9437
## Grid Depth 
## [1] 25
## Index of Optimal Lambda 
## [1] 9
## In-Sample MSFE
## [1] 1.881
## BigVAR Out of Sample MSFE
## [1] 4.552
## *** Benchmark Results *** 
## Conditional Mean Out of Sample MSFE
## [1] 5.285
## AIC Out of Sample MSFE
## [1] 4.879
## BIC Out of Sample MSFE
## [1] 5.167
## RW Out of Sample MSFE
## [1] 6.582
\end{verbatim}
\end{kframe}
\end{knitrout}

{\tt Model1Results} also includes
\begin{knitrout}
\definecolor{shadecolor}{rgb}{0.969, 0.969, 0.969}\color{fgcolor}\begin{kframe}
\begin{alltt}
\hlcom{# Coefficient matrix at end of evaluation period}
\hlstd{Model1Results}\hlopt{@}\hlkwc{betaPred}
\hlcom{# Residuals at end of evaluation period}
\hlstd{Model1Results}\hlopt{@}\hlkwc{resids}
\hlcom{# Lagged Values at end of evaluation period}
\hlstd{Model1Results}\hlopt{@}\hlkwc{Zvals}
\end{alltt}
\end{kframe}
\end{knitrout}

\subsection{Diagnostics and Additional Features}
\label{secDiag}
This section details the features of BigVAR that both tailor to the specific forecasting scenarios of the end-user and ensure that the most accurate possible forecasts are delivered.

\subsubsection{The ``Minnesota'' Lasso}
\label{secMN}
As opposed to shrinking every coefficient toward zero, all of the procedures in \BigVAR can be modified to instead shrink toward a vector random walk (i.e. $\PhiB^{(1)}=I_k$, all other coefficient matrices are still shrunk toward zero).  Such a modification is akin to the Bayesian VAR with Minnesota Prior of \cite{Litterman1979}.  This approach can be useful in scenarios exhibiting evidence of unit-root nonstationarity, which is commonplace in macroeconomic data.  For more details about this approach, see Section 4 of \cite{NicholsonSR}.

\BigVAR also allows for the option of shrinking some series toward zero while shrinking others toward a random walk.  This can be of use in applications, such as that presented in \cite{BGR}, in which a large cross section of series are examined; most are roughly stationary, but a few exhibit a substantial degree of persistence.

In examining Figure \ref{fig:macroplots}, we observe substantial persistence in the M1 series while the other series appear stationary.  We could attempt to shrink M1 toward a random walk while shrinking the others toward zero.  However, as can be observed below, doing so degrades forecast performance.   

\begin{knitrout}
\definecolor{shadecolor}{rgb}{0.969, 0.969, 0.969}\color{fgcolor}\begin{kframe}
\begin{alltt}
\hlstd{Model1MN} \hlkwb{<-} \hlkwd{constructModel}\hlstd{(}\hlkwd{as.matrix}\hlstd{(Y),} \hlnum{4}\hlstd{,} \hlstr{"HVARELEM"}\hlstd{,} \hlkwd{c}\hlstd{(}\hlnum{25}\hlstd{,} \hlnum{10}\hlstd{),} \hlkwc{T1} \hlstd{= T1,}
    \hlkwc{T2} \hlstd{= T2,} \hlkwc{verbose} \hlstd{=} \hlnum{FALSE}\hlstd{,} \hlkwc{MN} \hlstd{=} \hlnum{TRUE}\hlstd{,} \hlkwc{C} \hlstd{=} \hlkwd{c}\hlstd{(}\hlnum{0}\hlstd{,} \hlnum{0}\hlstd{,} \hlnum{0}\hlstd{,} \hlnum{1}\hlstd{))}
\hlstd{Model1MNresults} \hlkwb{<-} \hlkwd{cv.BigVAR}\hlstd{(Model1MN)}
\hlkwd{mean}\hlstd{(Model1MNresults}\hlopt{@}\hlkwc{OOSMSFE}\hlstd{)}
\end{alltt}
\begin{verbatim}
## [1] 4.569441
\end{verbatim}
\end{kframe}
\end{knitrout}

\subsubsection{Evaluating a Choice of Structure}

If the practitioner is unsure as to the choice of a VARX-L structure, one potential selection approach involves fitting a Sparse Lag or Sparse Own/Other VARX-L with both $\lambda$ and $\alpha$ selected by sequential cross validation.  The selected choice of $\alpha$ should provide some insight as to the importance of structure in the data.
\begin{knitrout}
\definecolor{shadecolor}{rgb}{0.969, 0.969, 0.969}\color{fgcolor}\begin{kframe}
\begin{alltt}
\hlcom{# Construct grid of candidate alphas between zero and 1}
\hlstd{alpha} \hlkwb{<-} \hlkwd{seq}\hlstd{(}\hlnum{0}\hlstd{,} \hlnum{1}\hlstd{,} \hlkwc{length} \hlstd{=} \hlnum{10}\hlstd{)}
\hlstd{Model2} \hlkwb{=} \hlkwd{constructModel}\hlstd{(}\hlkwd{as.matrix}\hlstd{(Y),} \hlkwc{p} \hlstd{=} \hlnum{4}\hlstd{,} \hlkwc{struct} \hlstd{=} \hlstr{"SparseLag"}\hlstd{,} \hlkwc{gran} \hlstd{=} \hlkwd{c}\hlstd{(}\hlnum{25}\hlstd{,}
    \hlnum{10}\hlstd{),} \hlkwc{verbose} \hlstd{=} \hlnum{FALSE}\hlstd{,} \hlkwc{VARX} \hlstd{=} \hlkwd{list}\hlstd{(),} \hlkwc{alpha} \hlstd{= alpha,} \hlkwc{T1} \hlstd{= T1,} \hlkwc{T2} \hlstd{= T2)}
\hlstd{SparseLagDiag} \hlkwb{<-} \hlkwd{cv.BigVAR}\hlstd{(Model2)}
\hlcom{# Selected value of alpha}
\hlstd{SparseLagDiag}\hlopt{@}\hlkwc{alpha}
\end{alltt}
\begin{verbatim}
## [1] 0.2222222
\end{verbatim}
\begin{alltt}
\hlcom{# Resulting out of sample MSFE}
\hlkwd{mean}\hlstd{(SparseLagDiag}\hlopt{@}\hlkwc{OOSMSFE}\hlstd{)}
\end{alltt}
\begin{verbatim}
## [1] 4.713579
\end{verbatim}
\begin{alltt}
\hlstd{Model3} \hlkwb{=} \hlkwd{constructModel}\hlstd{(}\hlkwd{as.matrix}\hlstd{(Y),} \hlkwc{p} \hlstd{=} \hlnum{4}\hlstd{,} \hlkwc{struct} \hlstd{=} \hlstr{"SparseOO"}\hlstd{,} \hlkwc{gran} \hlstd{=} \hlkwd{c}\hlstd{(}\hlnum{25}\hlstd{,}
    \hlnum{10}\hlstd{),} \hlkwc{verbose} \hlstd{=} \hlnum{FALSE}\hlstd{,} \hlkwc{VARX} \hlstd{=} \hlkwd{list}\hlstd{(),} \hlkwc{alpha} \hlstd{= alpha,} \hlkwc{T1} \hlstd{= T1,} \hlkwc{T2} \hlstd{= T2)}
\hlstd{SparseOODiag} \hlkwb{<-} \hlkwd{cv.BigVAR}\hlstd{(Model3)}
\hlcom{# Selected value for alpha}
\hlstd{SparseOODiag}\hlopt{@}\hlkwc{alpha}
\end{alltt}
\begin{verbatim}
## [1] 0.3333333
\end{verbatim}
\begin{alltt}
\hlcom{# Resulting out of sample MSFE}
\hlkwd{mean}\hlstd{(SparseOODiag}\hlopt{@}\hlkwc{OOSMSFE}\hlstd{)}
\end{alltt}
\begin{verbatim}
## [1] 4.674584
\end{verbatim}
\end{kframe}
\end{knitrout}
We observe that in the Sparse Lag setting, the selected value of $\alpha\approx 0.22$ is very close to our heuristic ($\frac{1}{k+1}=0.2$), indicating that the level of within-group sparsity determined by our heuristic is appropriate for this application.  In the Sparse Own/Other setting, the selected value is $\approx{0.33}$, indicating that a slightly greater degree of within-group sparsity than that imposed by the heuristic may be appropriate.

\subsubsection{Penalty Grid Position}

The {\tt plot} method of a {\tt BigVAR.results} object visualizes the position of $\hat{\lambda}$ over the grid of candidate values.  Figure \ref{fig:FIGLP} plots the in-sample MSFE for each value of $\lambda$ over the training period with the minimum value highlighted.  It is desirable for $\hat{\lambda}$ to be near the middle of the grid; if it is at the lower boundary, increasing the depth of the grid may lead to improved forecasting performance.  In examining Figure \ref{fig:FIGLP}, we see that $\hat{\lambda}$ is not at the lower boundary of the penalty grid.  If it were, we could simply construct a deeper penalty grid by increasing the first parameter of the ``gran'' argument in {\tt ConstructModel}.

\begin{knitrout}
\definecolor{shadecolor}{rgb}{0.969, 0.969, 0.969}\color{fgcolor}\begin{kframe}
\begin{alltt}
\hlkwd{plot}\hlstd{(Model1Results)}
\end{alltt}
\end{kframe}\begin{figure}

{\centering \includegraphics[width=\maxwidth]{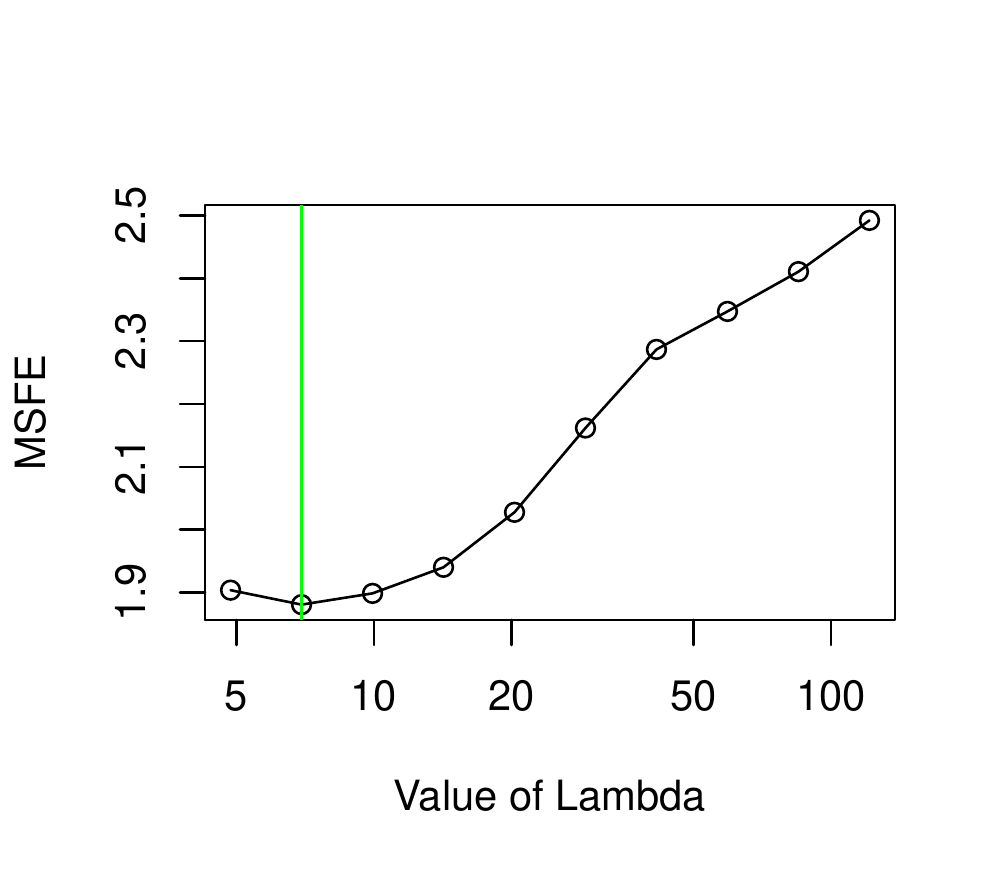} 

}

\caption[In-sample MSFE for each candidate penalty parameter]{In-sample MSFE for each candidate penalty parameter.}\label{fig:FIGLP}
\end{figure}

\end{knitrout}




\subsubsection{Visualizing Sparsity Patterns }

The method {\tt SparsityPlot.BigVAR.results} allows for the ability to view the sparsity pattern of the final estimated coefficient matrix $[\PhiB,\betaB]$ in the out of sample forecast evaluation period, which fits a model with the selected penalty parameter using all available data.  Figure \ref{fig:sparseplot} depicts this sparsity pattern for the Elementwise $\text{HVAR}_{4}(4)$ example.  Darker shading indicates coefficients that are larger in magnitude.  We observe that coefficients on the diagonal are larger in magnitude, indicating that ``own'' lags are relatively more important in forecasting than those of ``other'' series even though we did not explicitly consider an Own/Other structured grouping. 
\begin{knitrout}
\definecolor{shadecolor}{rgb}{0.969, 0.969, 0.969}\color{fgcolor}\begin{kframe}
\begin{alltt}
\hlkwd{SparsityPlot.BigVAR.results}\hlstd{(Model1Results)}
\end{alltt}
\end{kframe}\begin{figure}

{\centering \includegraphics[width=\maxwidth]{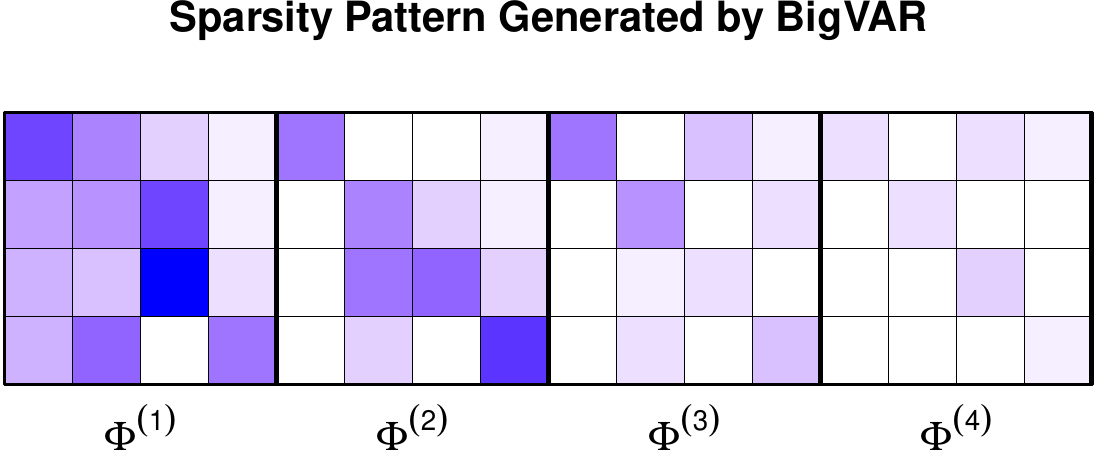} 

}

\caption[Sparsity plot generated by the Elementwise HVAR with active elements shaded]{Sparsity plot generated by the Elementwise HVAR with active elements shaded.  Darker coefficients are larger in magnitude.}\label{fig:sparseplot}
\end{figure}

\end{knitrout}

\subsubsection*{VARX-L Estimation}
If the user wishes to fit a VARX-L model, first the series should be arranged in a $T\times (k+m)$ matrix or such that the first $k$ columns are endogenous (modeled) series and the remaining $m$ are exogenous (unmodeled) series.

After doing so, a list of VARX specifications needs to be passed to {\tt constructModel}.  The list must contain two elements: $k$ denotes the number of endogenous series and $s$ the maximum lag order for the exogenous series.  For example, if we want to forecast GDP and the Federal Funds Rate using CPI and M1 as exogenous series (with $s=4$), we simply need to specify:  
\begin{knitrout}
\definecolor{shadecolor}{rgb}{0.969, 0.969, 0.969}\color{fgcolor}\begin{kframe}
\begin{alltt}
\hlstd{VARX} \hlkwb{=} \hlkwd{list}\hlstd{()}
\hlstd{VARX}\hlopt{$}\hlstd{k} \hlkwb{=} \hlnum{2}  \hlcom{# 2 endogenous series}
\hlstd{VARX}\hlopt{$}\hlstd{s} \hlkwb{=} \hlnum{4}  \hlcom{# maximum lag order of 4 for exogenous series}
\hlstd{Model2} \hlkwb{<-} \hlkwd{constructModel}\hlstd{(}\hlkwd{as.matrix}\hlstd{(Y),} \hlnum{4}\hlstd{,} \hlstr{"SparseLag"}\hlstd{,} \hlkwc{gran} \hlstd{=} \hlkwd{c}\hlstd{(}\hlnum{50}\hlstd{,} \hlnum{10}\hlstd{),} \hlkwc{VARX} \hlstd{= VARX,}
    \hlkwc{verbose} \hlstd{=} \hlnum{FALSE}\hlstd{)}
\hlstd{Model2Results} \hlkwb{=} \hlkwd{cv.BigVAR}\hlstd{(Model2)}
\end{alltt}
\end{kframe}
\end{knitrout}

\subsubsection*{N-step Ahead Out-of-sample Predictions}

Out of sample predictions can be obtained via the {\tt predict} method.  Multi-step ahead VAR forecasts are computed recursively using standard methods described in chapter 2 of \cite{Lutk}.  While $n$-step ahead predictions are available for VAR models, we currently only allow $1$-step ahead predictions for VARX models unless new data is provided.
\begin{knitrout}
\definecolor{shadecolor}{rgb}{0.969, 0.969, 0.969}\color{fgcolor}\begin{kframe}
\begin{alltt}
\hlcom{# One-step ahead VAR forecasts}
\hlkwd{predict}\hlstd{(Model1Results,} \hlnum{1}\hlstd{)}
\end{alltt}
\begin{verbatim}
##            [,1]
## [1,] -0.7786404
## [2,] -0.1048498
## [3,] -0.3420622
## [4,]  0.0893323
\end{verbatim}
\begin{alltt}
\hlcom{# Multi-step VARX prediction with new data}
\hlstd{VARXExample} \hlkwb{<-} \hlkwd{constructModel}\hlstd{(}\hlkwd{as.matrix}\hlstd{(Y),} \hlnum{4}\hlstd{,} \hlstr{"Basic"}\hlstd{,} \hlkwc{gran} \hlstd{=} \hlkwd{c}\hlstd{(}\hlnum{50}\hlstd{,} \hlnum{10}\hlstd{),} \hlkwc{VARX} \hlstd{=} \hlkwd{list}\hlstd{(}\hlkwc{k} \hlstd{=} \hlnum{2}\hlstd{,}
    \hlkwc{s} \hlstd{=} \hlnum{4}\hlstd{),} \hlkwc{verbose} \hlstd{=} \hlnum{FALSE}\hlstd{)}
\hlstd{result} \hlkwb{<-} \hlkwd{cv.BigVAR}\hlstd{(VARXExample)}
\hlcom{# Holdout data}
\hlstd{holdout}
\end{alltt}
\begin{verbatim}
##                CPI        FFR        GDP          M1
## 2015 Q3 -1.1046511 0.01641702 -0.8059263 -0.45622745
## 2015 Q4 -0.9337567 0.09110189 -1.1732646  0.05243381
## 2016 Q1 -1.0098161 0.10769852 -1.3022239  0.69904163
## 2016 Q2 -0.2235564 0.02471534 -0.6769686  0.93039422
## 2016 Q3 -0.5353280 0.02471534 -0.3663660  0.63849347
## 2016 Q4 -0.2087450 0.12429516 -0.6024822 -0.79893333
\end{verbatim}
\begin{alltt}
\hlkwd{predict}\hlstd{(result,} \hlkwc{n.ahead} \hlstd{=} \hlnum{3}\hlstd{,} \hlkwc{newxreg} \hlstd{=} \hlkwd{matrix}\hlstd{(holdout[,} \hlnum{3}\hlopt{:}\hlnum{4}\hlstd{],} \hlkwc{ncol} \hlstd{=} \hlnum{2}\hlstd{))}
\end{alltt}
\begin{verbatim}
##            [,1]         [,2]
## [1,] -0.2373474 -0.004659907
\end{verbatim}
\end{kframe}
\end{knitrout}

\subsubsection{Estimation with Fixed $\lambda$ }
\label{BVEST}
A user may wish to initially estimate $\hat{\lambda}$ by rolling cross-validation and continue to use that value as new data becomes available or potentially apply their own penalty parameter selection technique.  In such a scenario, it would not be desirable to fit a model with {\tt cv.BigVAR}.

We provide an alternative function {\tt BigVAR.est}, which requires an object of class \BigVAR as input and fits a VARX-L or HVAR model using all available data for either a fixed grid of $\lambda$ values or a grid determined by the data (as is done in {\tt cv.BigVAR}).  For example, suppose that wish to re-estimate our Elementwise $\text{HVAR}_{4}(4)$ model with newly available data using the $\hat{\lambda}$ that was selected in Section \ref{IMP1}.       

\begin{knitrout}
\definecolor{shadecolor}{rgb}{0.969, 0.969, 0.969}\color{fgcolor}\begin{kframe}
\begin{alltt}
\hlcom{# new data}
\hlstd{holdout}
\hlcom{# augment data in original BigVAR object}
\hlstd{Model1}\hlopt{@}\hlkwc{Data} \hlkwb{<-} \hlkwd{as.matrix}\hlstd{(}\hlkwd{rbind}\hlstd{(Y, holdout))}
\hlcom{# Extract the optimal lambda from our BigVAR results object}
\hlstd{lambda} \hlkwb{<-} \hlstd{Model1Results}\hlopt{@}\hlkwc{OptimalLambda}
\hlcom{# Set ownlambas indicator TRUE in BigVAR object}
\hlstd{Model1}\hlopt{@}\hlkwc{ownlambdas} \hlkwb{=} \hlnum{TRUE}
\hlcom{# Replace granularity specs with choice of lambda}
\hlstd{Model1}\hlopt{@}\hlkwc{Granularity} \hlkwb{<-} \hlstd{lambda}
\hlkwd{BigVAR.est}\hlstd{(Model1)}
\hlcom{# returns a list containing: a k x (kp+ms+1) x n array, in which n denotes}
\hlcom{# the number of penalty parameters a vector of penalty parameters}
\hlcom{# corresponding to each slice of the array}
\end{alltt}
\end{kframe}
\end{knitrout}


\subsubsection{Simulating multivariate time series}
When developing new methods, it is often good practice to evaluate their performance on simulated data.  \BigVAR offers the ability to simulate realizations from user-provided VAR coefficient and covariance matrices via the function {\tt MultVARSim}.  In order to simulate from a $\text{VAR}_k(p)$, we convert its coefficient matrix to block companion form (following equation 2.1.8 in \cite{Lutk}):   
\begin{align}
\label{bigcm}
\mathbf{A}=
\begin{bmatrix}
\PhiB^{(1)} & \PhiB^{(2)} & \dots & \PhiB^{(p-1)} & \PhiB^{(p)}\\
\utwi{I}_{k}& \utwi{0} &\utwi{0} &\utwi{0}&\utwi{0}\\
\utwi{0}& \utwi{I}_{k} &\utwi{0} &\utwi{0}&\utwi{0}\\
\vdots & \vdots & \ddots & \vdots &\vdots\\ 
\utwi{0}& \utwi{0} &\utwi{0} &\utwi{I}_k&\utwi{0}
\end{bmatrix}
\end{align}

An example is shown below.

\begin{knitrout}
\definecolor{shadecolor}{rgb}{0.969, 0.969, 0.969}\color{fgcolor}\begin{kframe}
\begin{alltt}
\hlcom{# included VAR_3(3) coefficient matrix in BigVAR in block companion form}
\hlkwd{data}\hlstd{(Generator)}
\hlstd{k} \hlkwb{<-} \hlnum{3}
\hlstd{A[}\hlnum{1}\hlopt{:}\hlstd{k,]}
\end{alltt}
\begin{verbatim}
##       [,1]  [,2] [,3]  [,4]  [,5]  [,6]  [,7]  [,8] [,9]
## [1,] -0.29  0.00  0.0 -0.62  0.00  0.00 -0.49  0.00 0.00
## [2,] -0.26 -0.20  0.0 -0.77 -0.36  0.00 -1.24 -0.07 0.00
## [3,] -0.66  0.75  1.3  0.30 -0.40 -0.44  0.36  0.05 0.03
\end{verbatim}
\begin{alltt}
\hlstd{SigmaU} \hlkwb{<-} \hlnum{.01}\hlopt{*}\hlkwd{diag}\hlstd{(k)} \hlcom{#Scaled identity covariance}
\hlstd{YSim} \hlkwb{<-} \hlkwd{MultVARSim}\hlstd{(k,A,}\hlnum{3}\hlstd{,SigmaU,}\hlkwc{T}\hlstd{=}\hlnum{100}\hlstd{)}
\end{alltt}
\end{kframe}
\end{knitrout}

When constructing a coefficient matrix, one needs to be judicious in ensuring stationarity.  Stationarity requires that the all eigenvalues of $\mathbf{A}$ have modulus less than 1.  As stated in \cite{McElroy}, there is generally no link between the magnitude of elements in a coefficient matrix and stationarity.  For example, consider the case where $k=2$ and $p=1$.  The $\text{VAR}_{2}(1)$ coefficient matrix
\begin{align}
\label{bigcm}
\PhiB=
\begin{bmatrix}
0 & 0\\
\epsilon & 0
\end{bmatrix}
\end{align}
is stationary for any value of $\epsilon$. 

Recent developments by \cite{bosh} provide a framework to guarantee stationary VAR coefficient matrices, but their method cannot impose structured sparsity, which limits its utility in evaluating the performance of the VARX-L and HVAR class of models.                   


\subsection{Structural Macroeconomic Analysis}
\label{secIRFM}
Though \BigVAR is primarily designed to forecast high-dimensional time series, it can also be of use in analyzing the joint dynamics of a group of interrelated time series.  In order to conduct policy analysis, many macroeconomists make use of VARs to examine the impact of shocks to certain variables on the entire system (holding all other variables fixed).  This is know as impulse response analysis.  It has the potential to be very important in a high-dimensional setting as omitting variables from a system can lead to major distortions \citep{Lin}.

For example, a macroeconomist may wish to analyze the impact of a 100 basis point increase in the Federal Funds Rate on all included series over the next 8 quarters.  To do so, we can utilize the function {\tt generateIRF}, which converts the last estimated coefficient matrix to fundamental form (for details, see Section \ref{secIRF} in the appendix).  The impulse responses generated from this ``shock'' are depicted in Figure \ref{fig:FIGIRF}.  
\begin{knitrout}
\definecolor{shadecolor}{rgb}{0.969, 0.969, 0.969}\color{fgcolor}\begin{figure}[h]

{\centering \includegraphics[width=\maxwidth]{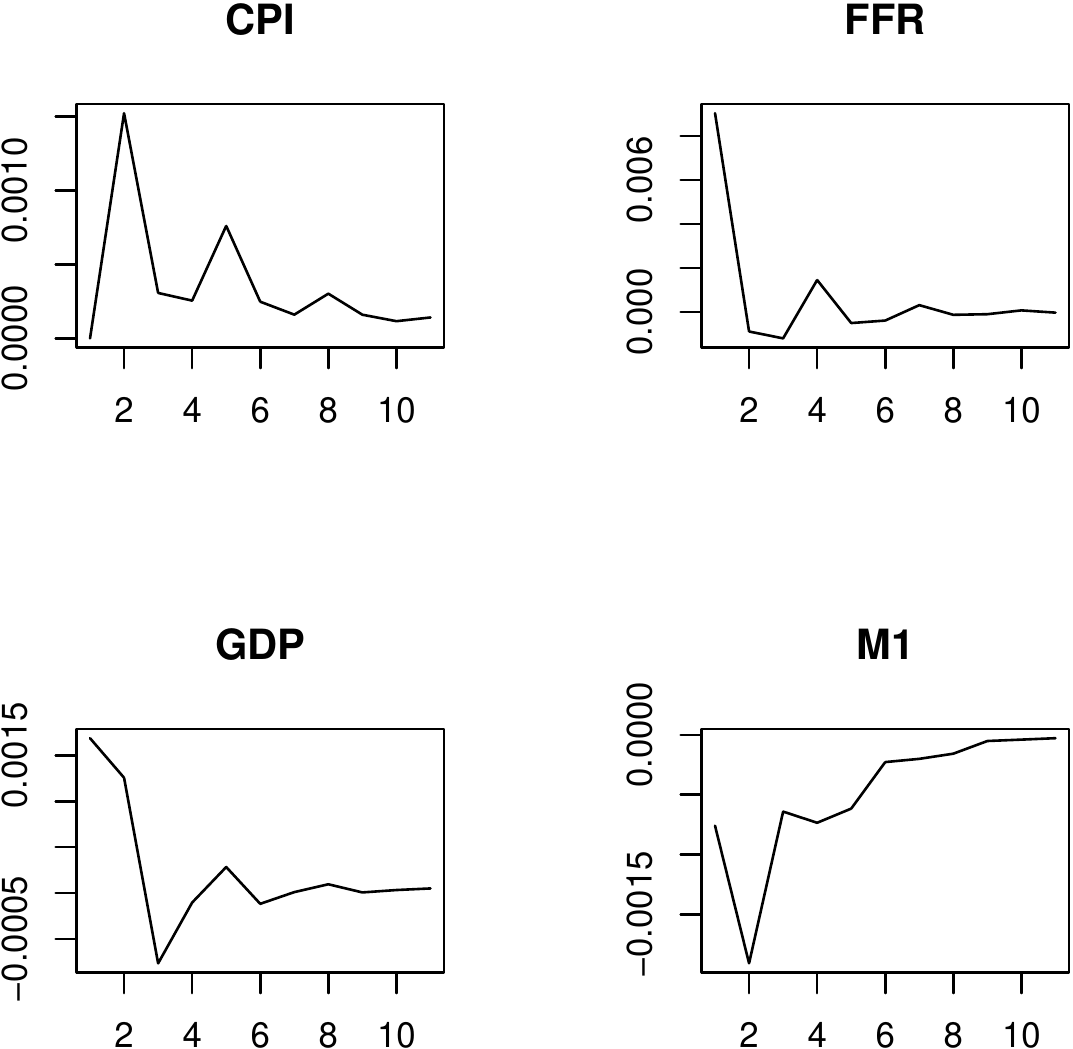} 

}

\caption[Impulse responses generated as the result of a 100 basis point increase to the Federal Funds Rate]{Impulse responses generated as the result of a 100 basis point increase to the Federal Funds Rate}\label{fig:FIGIRF}
\end{figure}

\end{knitrout}

\subsection{Information Criterion Benchmarks}
By default, we compare our methods to the conventional approach of selecting from a universal, sequentially increasing lag order as chosen by AIC or BIC and fitting the resulting VAR or VARX by least squares. Due to poor numerical stability as well as substantial computational overhead, we do not recommend including this benchmark when working in high dimensions (i.e. $kp\approx T)$, hence we offer the option to disable it (by setting IC to {\tt FALSE} in {\tt constructModel}).  

We implement the numerically stable and computationally efficient technique proposed in \cite{Neumaier}, which calculates the least squares VARX using a QR decomposition that does not require explicit matrix inversion. \BigVAR contains two functions that fit least squares VAR and VARX according to information criterion minimization.  {\tt VARXForecastEval}, which evaluates the $h$-step ahead forecasting performance of a VAR or VARX with lags selected by AIC or BIC over an evaluation period, and {\tt VARXFit}, which fits a least squares VAR or VARX with the lag order selected by AIC or BIC.  {\tt VARXForecastEval} is called automatically by {\tt cv.BigVAR} if {\tt IC} is set to {\tt TRUE} in {\tt constructModel}.  Implementation details are provided in Section \ref{secIC} of the appendix.

\begin{knitrout}
\definecolor{shadecolor}{rgb}{0.969, 0.969, 0.969}\color{fgcolor}\begin{kframe}
\begin{alltt}
\hlcom{# Least Squares AIC VARX}
\hlstd{LSAIC} \hlkwb{<-} \hlkwd{VARXFit}\hlstd{(Y,} \hlnum{12}\hlstd{,} \hlstr{"AIC"}\hlstd{,} \hlkwa{NULL}\hlstd{)}
\hlcom{# VARX Forecast Eval with BIC Pass in matrix of zeros for exogenous series}
\hlcom{# This matrix is not used in the VAR setting}
\hlstd{X} \hlkwb{<-} \hlkwd{matrix}\hlstd{(}\hlnum{0}\hlstd{,} \hlkwc{nrow} \hlstd{=} \hlkwd{nrow}\hlstd{(Y),} \hlkwc{ncol} \hlstd{=} \hlnum{1}\hlstd{)}
\hlcom{# Shift by p quarters to account for initialization in order to match}
\hlcom{# cv.BigVAR output}
\hlstd{BICEval} \hlkwb{<-} \hlkwd{VARXForecastEval}\hlstd{(}\hlkwd{as.matrix}\hlstd{(Y)[(p} \hlopt{+} \hlnum{1}\hlstd{)}\hlopt{:}\hlkwd{nrow}\hlstd{(Y), ], X, p,} \hlnum{0}\hlstd{, T2} \hlopt{-} \hlstd{p,}
    \hlkwd{nrow}\hlstd{(Y)} \hlopt{-} \hlstd{p,} \hlstr{"BIC"}\hlstd{,} \hlnum{1}\hlstd{)}
\hlkwd{mean}\hlstd{(BICEval)}
\end{alltt}
\begin{verbatim}
## [1] 5.18401
\end{verbatim}
\end{kframe}
\end{knitrout}

\section{Refitting with least squares}
\label{sec4}
Within the regularization framework, it is often of interest to use the lasso and its structured variants for variable selection, while refitting the support selected according to least squares.  \cite{cherno1} prove that a post-selection least squares refitting procedure has smaller bias than the conventional lasso in the univariate regression setting.  In this section, we extend the refitting framework to the VAR context and perform a detailed simulation study to explore the forecasting performance of several potential refitting procedures. 

Post-selection estimation has been considered in time-dependent problems by \cite{BickelSong}, who refit by least squares based upon the support chosen by their structured VAR penalties.  However, such an approach does not take into account $\Sigma_u,$ the VAR innovation covariance matrix.  A seminal result from \cite{Zellner} shows that, in the absence of parameter restrictions, ordinary least squares and generalized least squares coincide in the VAR framework.  However, once parameter restrictions are introduced, generalized least squares is more efficient.

A feasible generalized least squares VAR that incorporates parameter restrictions (such as setting coefficients to zero) is introduced in \cite{brug} and is utilized by \cite{Davis} in the context of constrained maximum likelihood VAR estimation.  Details of this approach are provided by Equations \eqref{SigmaUEst} and \eqref{FGLS} in Section \ref{secRVAR} of the appendix.

We have found this formulation to be unsuited for our framework.  First, in the early stages of rolling cross validation for short series, we often face scenarios in which the number of potential least squares parameters is close to exceeds the length of the series.  Hence, taking the inverse of a poorly conditioned covariance matrix in these situations results in substantial estimation error.  As an alternative, we propose extending the iterated feasible generalized least squares approach developed by \cite{foschithesis}, which formulates the feasible generalized least squares problem in a framework that avoids explicit matrix inversion.  Details of our implementation are provided in Section \ref{secifgls} in the appendix.

\subsection{Simulation Study}

In this section, we conduct a detailed simulation study to evaluate the forecasting performance of several refitting procedures.  First, we consider the conventional \emph{relaxed least squares} approach which simply refits the support selected according to restricted least squares (as defined by Equation \eqref{RLS} in the appendix).  Second, we consider a \emph{weighted relaxed least squares approach} which refits according to feasible generalized least squares using a covariance matrix with the diagonal entries set to the unconditional variance of each marginal series, and all other elements set to zero.  Next, we consider the iterated feasible GLS approach, which iteratively refines the covariance matrix utilizing the procedure outlined in Algorithm \ref{algIC} the appendix.  Finally, we compare against the ``oracle'' procedure in which we perform generalized least squares using the covariance matrix from which the data was generated.

In this section, we operate exclusively in the Basic VAR-L setting.  We do not believe that it is appropriate to refit when imposing structure; the groupings impose a ridge-like regularization effect which is not preserved after a least squares transformation.  

We consider simulating from a $\text{VAR}_{8}(4)$ with an unstructured sparsity pattern as depicted in Figure \ref{fig:sparseplotsim} and we consider four covariance matrices that are discussed in the following sections.  For each covariance matrix, we simulate a VAR of length 200 and use the middle third of the data for penalty parameter selection and the final third for forecast evaluation.  We record the average 1-step ahead MSFE over the evaluation period for each simulation and repeat this process 100 times.

\begin{knitrout}
\definecolor{shadecolor}{rgb}{0.969, 0.969, 0.969}\color{fgcolor}\begin{figure}

{\centering \includegraphics[width=\maxwidth]{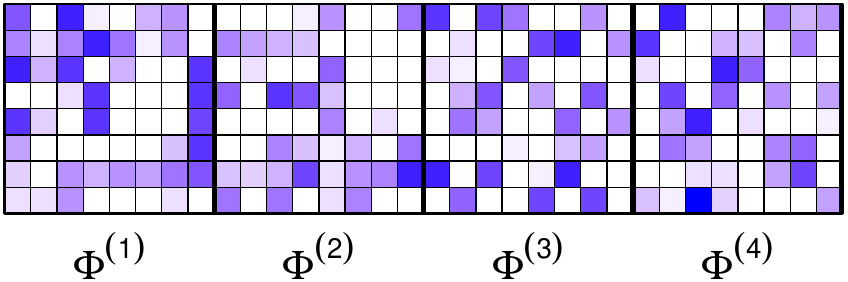} 

}

\caption[Sparsity Pattern of the $VAR_8(4) $ Coefficient Matrix Used in all Simulation Scenarios]{Sparsity Pattern of the $VAR_8(4) $ Coefficient Matrix Used in all Simulation Scenarios}\label{fig:sparseplotsim}
\end{figure}

\end{knitrout}
\subsubsection*{Simulation Scenario 1: Sparse Hub Structure}

The first covariance matrix we consider is sparse with two \emph{cliques}.  Each series within the clique has identical covariance and it is set to zero outside of the clique.  The variance is identical across all observations.  Note that since we do not impose sparsity in our covariance estimation, our IFGLS procedure will not be able to capture this structure.  

\begin{knitrout}
\definecolor{shadecolor}{rgb}{0.969, 0.969, 0.969}\color{fgcolor}\begin{figure}

{\centering \includegraphics[width=\maxwidth]{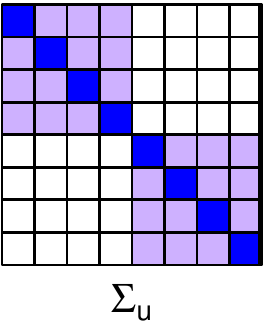} 

}

\caption[Covariance Matrix Used in Simulation Scenario 1]{Covariance Matrix Used in Simulation Scenario 1}\label{fig:sigma1plot}
\end{figure}

\end{knitrout}

  \begin{table}[h]
    \centering    
 \caption{Out of sample MSFE of one-step ahead forecasts after 100 simulations: Scenario 1}
\label{tab:tabresRV}
   \begin{tabular}{ l c c }
Model & Average MSFE & Standard Error \\
    \hline
Basic  & 2.4527 & 0.0188\\
Basic Relaxed Least Squares & 2.4756 & 0.0201 \\
Basic Weighted Least Squares & 2.4991 & 0.0204\\
Basic IFGLS & 2.4526 & 0.0199\\
Basic Oracle & 2.4483 & 0.0197\\
\hline
Sample Mean & 3.6568 & 0.0370\\
Random Walk & 7.0373 & 0.0808\\
Least Squares AIC VAR & 2.9113 & 0.0236\\
Least Squares BIC VAR & 3.6212 & 0.0357\\
  \end{tabular}
 \end{table}

 We observe that the Oracle GLS achieves the best forecasting performance, though both the IFGLS procedure and the Basic VAR-L are well within one standard error.  The relaxed least squares outperforms weighted least squares which subsequently outperforms all naive methods.  It should be noted that all Basic VAR-L methods achieve very similar forecast performance, suggesting that in this scenario, there is little to be gained in terms of forecasting improvements by refitting. 
 
 
\subsubsection{Simulation Scenario 2: Poorly Conditioned} 

We next consider simulating using a covariance matrix with a high condition number.  We constructed the covariance matrix to have a condition number of 50,214,428.  In such a scenario, the conventional feasible GLS estimator is inadvisable as computing $\Sigma_u^{-1}$ will result in substantial estimation error.  This is a scenario that we have encountered in the early stages of sequential cross validation, in which the length of the time series is relatively small compared to the number of potential model coefficients.  Under this scenario, we should expect the Oracle GLS estimator to perform very poorly as a result of this imprecision.  Since our IFGLS procedure does not require explicit matrix inversion, it should be relatively robust to a poorly conditioned covariance matrix. 

\begin{knitrout}
\definecolor{shadecolor}{rgb}{0.969, 0.969, 0.969}\color{fgcolor}\begin{figure}

{\centering \includegraphics[width=\maxwidth]{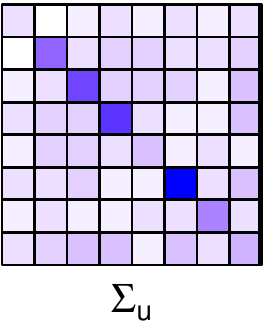} 

}

\caption[Covariance Matrix Used in Simulation Scenario 2]{Covariance Matrix Used in Simulation Scenario 2}\label{fig:sigma2plot}
\end{figure}

\end{knitrout}

  \begin{table}[h]
    \centering    
 \caption{Out of sample MSFE of one-step ahead forecasts after 100 simulations: Scenario 2}
\label{tab:tabresRV}
   \begin{tabular}{ l c c }
Model & Average MSFE & Standard Error \\
    \hline
Basic  & 6.4045 & 0.0519\\
Basic Relaxed Least Squares & 6.5135 & 0.0539 \\
Basic Weighted Least Squares & 6.6003 & 0.0588\\
Basic IFGLS & 6.4464 & 0.0544\\
Basic Oracle & 6.5979 & 0.0561\\
\hline
Sample Mean & 9.7013 & 0.1106\\
Random Walk & 18.5767 & 0.2505\\
Least Squares AIC VAR & 7.5102 & 0.0679\\
Least Squares BIC VAR & 7.5102 & 0.0679\\
  \end{tabular}
 \end{table}
 
 Under this scenario, we find that any form of refitting only serves to degrade forecast performance; the Basic VAR-L achieves the best performance.  The IFGLS performs relatively well, better than any other refitting method and within one standard error of the Basic VAR-L.  The  Oracle GLS, as it is trying to incorporate a nearly singular covariance, achieves relatively poor performance, on par with weighted least squares.  Notice that the AIC and BIC VARs, both of which incorporate the covariance in lag order selection, achieve the exact same forecasting performance.


\subsubsection*{Scenario 3: Scaled Identity}
We next consider the case in which the covariance matrix is set to $0.1\times I_k$.  This scenario examines the robustness of the IFGLS framework in cases where an estimate of the covariance should provide no aid in forecasting.  The results from this scenario are detailed in Table \ref{tab:tab3a}.

  \begin{table}[h]
    \centering    
 \caption{Out of sample MSFE of one-step ahead forecasts after 100 simulations: Scenario 3}
\label{tab:tab3a}
   \begin{tabular}{ l c c }
Model & Average MSFE & Standard Error \\
    \hline
Basic  & 0.8700 & 0.0055\\
Basic Relaxed Least Squares & 0.8779 & 0.0057  \\
Basic Weighted Least Squares & 0.8775 & 0.0060\\
Basic IFGLS & 0.8782 & 0.0060\\
Basic Oracle & 0.8773 & 0.0060\\
\hline
Sample Mean & 1.3218 & 0.0122\\
Random Walk & 2.6096 & 0.0285\\
Least Squares AIC VAR & 1.0273 & 0.0073\\
Least Squares BIC VAR & 1.3219 & 0.0122\\
  \end{tabular}
 \end{table}

 In this setting, we again find that the Basic VAR-L achieves the best forecasting performance, substantially outperforming all refitting procedures, which are all within one standard error of each other.  This suggests that in settings in which there is no contemporaneous dependence, any type of refitting will only serve to degrade forecast performance.

 \subsubsection*{Scenario 4: Dense matrix}

\begin{knitrout}
\definecolor{shadecolor}{rgb}{0.969, 0.969, 0.969}\color{fgcolor}\begin{figure}

{\centering \includegraphics[width=\maxwidth]{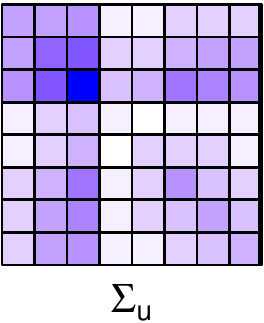} 

}

\caption[Covariance Matrix Used in Simulation Scenario 4]{Covariance Matrix Used in Simulation Scenario 4}\label{fig:figWC}
\end{figure}

\end{knitrout}
 
 Our final scenario considers a well-conditioned dense covariance matrix as shown in Figure \ref{fig:figWC}.  In this setting, we should expect the IFGLS estimator and Oracle to achieve the best performance, as they are best able to capture the true covariance structure.

  \begin{table}[h]
    \centering    
 \caption{Out of sample MSFE of one-step ahead forecasts after 100 simulations: Scenario 4}
\label{tab:tabres4}
   \begin{tabular}{ l c c }
Model & Average MSFE & Standard Error \\
    \hline
Basic Unrelaxed  & 3.0012 & 0.0446\\
Basic Relaxed Least Squares & 3.0355 & 0.0447  \\
Basic Weighted Least Squares & 3.0598 & 0.0448\\
Basic IFGLS & 2.9641 & 0.0406\\
Basic Oracle & 3.1561 & 0.0510\\
\hline
Sample Mean & 4.4773 & 0.0719\\
Random Walk & 7.9507 & 0.1340\\
Least Squares AIC VAR & 3.5527 & 0.0550\\
Least Squares BIC VAR & 3.5527 & 0.0550\\
  \end{tabular}
 \end{table}

 We find that the IFGLS procedure achieves the best forecasting performance, followed by the unrelaxed Basic VAR-L.  All other refitting procedures perform substantially worse.  Surprisingly, the Oracle GLS performs the worst of any regularization procedure.  This demonstrates yet again that even if we can obtain a reliable estimate for the covariance matrix, it provides no guarantee of forecasting improvement.



\subsubsection{Empirical Example}

We additionally consider examining the performance of these models on the macroeconomic data examined in Section \ref{IMP1}.

  \begin{table}[h]
    \centering    
 \caption{\footnotesize Out of sample MSFE of one-step ahead forecasts of 4 US macroeconomic series}
\label{tab:tabresRV}
   \begin{tabular}{ l c  }
Model &  MSFE over Evaluation Period  \\
    \hline
Basic Unrelaxed  & 4.7353 \\
Basic Relaxed Least Squares & 4.7995 \\
Basic Weighted Least Squares & 4.8196\\
Basic IFGLS & 4.8440 \\
\hline
Sample Mean & 5.3722 \\
Random Walk & 6.8677 \\
Least Squares AIC VAR & 5.3722 \\
Least Squares BIC VAR & 5.4281 \\
  \end{tabular}
 \end{table}

 In applying refitting procedures to actual data, we find that none of the proposed methods lead to forecasting improvements over the Basic VAR-L.

 \subsection{Summary}

 We observe that in most simulation scenarios as well as our empirical application, refitting does not lead to substantial improvements in forecasting performance.  This lack of improvement is likely due to several factors.  First, it is possible that a different penalty parameter selection procedure is more appropriate when refitting is involved.  In our experience, the penalty parameter selected by sequential cross validation tends to ``over-select'' model coefficients, choosing many coefficients that are technically active, but extremely small in magnitude.  It would appear that very small magnitude model coefficients should not be refit, but defining a cutoff magnitude is challenging.     
 
 \cite{cherno2} develop a hypothesis testing procedure that can be used to determine which coefficients to refit, but it does not extend to a multivariate time-dependent setting.  In addition, as pointed out by \cite{cherno1}, when refitting, the optimal penalty parameter should larger than in the unrelaxed setting.  This suggests that proper incorporation of refitting requires the development of an alternative penalty parameter selection procedure that encourages more sparse solutions.  
 
 Despite its relatively lackluster forecasting performance, the IFGLS framework could be potentially useful in applications other than forecasting, such as generating impulse response functions (as discussed in Section \ref{secIRFM}), in which a reliable estimate of the innovation covariance matrix is crucial for an accurate depiction of the joint dynamics of the included series.

\section{Conclusion}
\label{secconc}

\BigVAR offers a convenient framework for the forecasting of high-dimensional multivariate time series with structured convex penalties.  Our methodology is transparent and can easily be understood and applied by practitioners and academics alike.  Our package is currently available on the Comprehensive R Archive Network at \url{https://cran.r-project.org/web/packages/BigVAR/index.html} and the development version is hosted on GitHub \url{https://github.com/wbnicholson/BigVAR}.       


\bibliography{finalbib}

\appendix
\section{Appendix}

\subsection{Notation}
When detailing our algorithms, we find it convenient to express the VARX in compact matrix notation.
\begin{align*}
 &   \begin{array}[r]{llll}
       \Y=[\mathbf{y}_1,\dots,\mathbf{y}_T]& (k\times T);  & \quad \mathbf{X}=[\mathbf{x}_1,\dots,\mathbf{x}_T]& (m\times T).\\
\Z_t=
[1,\mathbf{y}_t^{\top},\dots,\mathbf{y}_{t-p}^{\top},
\mathbf{x}_{t}^{\top},\dots,
\mathbf{x}_{t-s}^{\top} 
] & [1 \times (kp+ms+1)];&\quad
\Z=
[\Z_2;\dots;\Z_{T-1}]
 & [T\times (kp+ms+1)]^{\top};\\
\PhiB=[\PhiB^{(1)},\PhiB^{(2)},\dots,\PhiB^{(p)}] & (k\times kp);& \quad \betaB=[\betaB^{(1)},\dots,\betaB^{(s)}]& [k\times ms];\\
  \B=
[\nu,\PhiB,\betaB] &
 [k\times (kp+ms+1)]
;&\quad \mathbf{U}=[\mathbf{u}_1,\dots,\mathbf{u}_T] & (k\times T)
\end{array}
\end{align*}
We can then express the VARX as   
\begin{align*}
 \Y=\B\Z+\mathbf{U},
\end{align*}
in which $\mathbf{U}\stackrel{\text{iid}}{\sim} (0,I_T\otimes \Sigma_u)$.

\subsection{Computing Information Criterion Based Benchmarks}
\label{secIC}
Following \cite{Neumaier}, we construct the matrix $\mathbf{K}=[\Z^{\top},\Y^{\top}]$.  We then compute a QR factorization
\begin{align*}
\mathbf{K}=QR,  
\end{align*}
in which Q is an orthogonal matrix and R is upper triangular of the form:
\begin{align*}
&R= \begin{blockarray}{c@{}cc@{\hspace{4pt}}cl}
    & kp+ms+1 & k  & & \\
    \begin{block}{[c@{\hspace{5pt}}cc@{\hspace{5pt}}c]l}
      & R_{11} & R_{12}  & & kp+ms+1 \\
      & 0 & R_{22}  & & k \\
    \end{block}
  \end{blockarray}  
\end{align*}
Then, we can compute the least squares estimate $\widehat{\B}$ as
\begin{align*}
\widehat{\B}&=\big(R_{12}^{\top}R_{11}(R_{11}^{\top}R_{11})^{-1}\big)^{\top},\\
&=\big(R_{12}^{\top}R_{11}R_{11}^{-1}(R_{11}^{\top})^{-1}\big)^{\top},\\
&=\big(R_{12}^{\top}(R_{11}^{\top})^{-1}\big)^{\top},\\
&=\big(R_{11}^{-1}R_{12}\big)^{\top},  
\end{align*}
which can be evaluated with a triangular solver, hence does not require explicit matrix inversion.  We can then obtain the residual covariance $\widehat{\Sigma}_u$ as:
\begin{align*}
\frac{R_{22}^{\top}R_{22}}{T}
\end{align*}

Our implementation of this procedure in the context of VAR and VARX lag order selection is described in Algorithm \ref{algIC} in Section \ref{tabalg}.

\subsection{Generating Impulse Response Functions}
\label{secIRF}
In order to perform impulse response analysis, the system needs to be identified, hence we need to convert the VAR to a moving average representation.  Following \cite{Lutk} and \cite{Lin}, we can convert a VAR(p) process to MA form as follows.  First convert the VAR(p) to VAR(1), as in Equation \eqref{bigcm}. 
\begin{align*}
\Y_t=\utwi{\nu}+\mathbf{A}\Y_{t-1}+\mathbf{U}_t.  
\end{align*}
 Then, (assuming the coefficient matrix generates a stationary process), it can be represented as
 \begin{align}
\label{MA1}
   \Y_t=\utwi{\mu}+\sum_{i=0}^{\infty}\mathbf{A}^{i}U_{t-i},
 \end{align}
We then obtain the MA representation by pre-multiplying both sides of Equation (\ref{MA1}) by the $k\times kp$ matrix J = $[I_k, \utwi{0},\utwi{0}]$, resulting in
\begin{align*}
&J\Y_t=J\utwi{\nu}+\sum_{i=0}^{\infty}J\mathbf{A}^{i}J^{\top}JU_{t-i},\\  
&\mathbf{y}_t=\utwi{\nu}+\sum_{i=0}^{\infty}\Gamma_{i}\mathbf{u}_{t-i},  
\end{align*}
in which $\Gamma_i$ is the MA coefficient matrix (constructed by $J\mathbf{A}^{i}J^{\top}$) measuring the impulse response.  Since, the covariance of $\mathbf{U}_t$ is not diagonal, we cannot perform impulse response without a factorization.  Traditionally, the Cholesky Decomposition is used to factor $\Sigma_u=CC^{\top}$, where $C$ is a lower triangular matrix.  After this factorization, the MA coefficients can be expressed as:
\begin{align}
  \label{IRF1}
 \Y_t=\sum_{i=0}^{\infty} \Theta_iw_{t-i}, 
\end{align}
in which $\Theta_i=\Gamma_iC$, $w_t=C^{-1}U_t$, $E(w_tw_t^{\top})=I_k$.  Then, with D representing the diagonal of the Cholesky factor P and defining $W=CD^{-1}, \Sigma_w=DD^{\top}$, it is possible to use Equation \eqref{IRF1} to model a response function to trace the effect of a shock over $n$ periods by examining $\Theta_i$ for $i=1,\dots,n$.

\subsection{Relaxed (Group) Lasso-VAR}
\label{secRVAR}
Since  the lasso and its structured counterparts are known to shrink non-zero regression coefficients, in practice, they are often used for model selection followed by refitting the reduced model using least squares \citep{meinsh}, \citep{cherno1}).  This approach has been extended to the VAR setting by \cite{BickelSong}, who briefly remark that they use their group lasso models for variable selection and refit based on least squares.

Refitting with least squares in the penalized VAR setting is inefficient and completely ignores the VAR's structure.  As demonstrated in \cite{Zellner}, in the absence of parameter restrictions, the ordinary and generalized least squares estimators (GLS) coincide in the VAR framework.  However, once restrictions are introduced, the generalized least squares estimator is asymptotically more efficient than ordinary least squares.

An estimation procedure that can take into account linear restrictions (such as fixing some parameters at zero) is referred to in the time series literature as a ``Restricted VAR,'' and was explored in the context of constrained likelihood Lasso-VAR estimation by \cite{Davis}.  As we use this method to re-estimate nonzero coefficients, to avoid confusion we will refer to this two-step estimation procedure as a ``Relaxed Basic VAR-L.''

\subsubsection{Notation}

Let $\widehat{\PhiB}$ denote the coefficient matrix obtained from a structured regularization procedure (e.g. a Basic VAR-L), that returned $r$ nonzero coefficients.  The selected coefficients can be expressed as linear constraints of the form
\begin{align}
\label{restr}
\text{vec}(\widehat{\PhiB})=\mathbf{R}\hat{\phi},
\end{align}
in which $\mathbf{R}$ is a $(k^2p+k)\times r$ selection matrix of rank $r$ consisting of columns from an identity matrix of dimension $k^2p+k$, and $\hat{\phi} = \text{vec}( \{ \widehat \PhiB : \widehat \PhiB_{jk} \ne 0 \})$.  Within the relaxed framework, $\lambda$ is held constant and the support recovered is taken as given.  Following \cite{brug},  we can express the GLS estimator of the Relaxed VAR as
\begin{align}
\label{rv1}
&\vect(\widehat{\PhiB}^{GLS})=\mathbf{R}[\mathbf{R}^{\top}(\Z\Z^{\top}\otimes  \Sigma_u^{-1})\mathbf{R}]^{-1}\mathbf{R}^{\top}(\Z \otimes \Sigma_u^{-1})\text{vec}(\Y),
\end{align}
in which $\otimes$ denotes the Kronecker product.  However, since $\Sigma_u$ is unknown in general, Equation (\ref{rv1}) cannot be used in practice.  The two step procedure to construct a ``feasible'' GLS estimator starts by calculating the Relaxed Least Squares (RLS) estimator
\begin{align}
\label{RLS}
\vect(\widehat{\PhiB}^{\text{Rlx}})=\mathbf{R}[\mathbf{R}^{\top}(\Z\Z^{\top}\otimes I_k)\mathbf{R}]^{-1}\mathbf{R}^{\top}(\Z \otimes I_k)\text{vec}(\Y).
\end{align}
The RLS estimator is then used to estimate $\Sigma_u$.  If estimating $\Sigma_u$ is not tractable, which can occur when the series length T is small relative to the number of component series $k$, $\hat{\PhiB}^{\text{Rlx}}$ can be used to return ``unshrunk'' parameter estimates under the assumption that $\Sigma_u$ is the identity matrix.  Otherwise $\Sigma_u$ can be estimated by
\begin{align}
\label{SigmaUEst}
 \widehat{\Sigma}_u=\frac{1}{T-(p\times k)}(\Y-\widehat{\PhiB}^{\text{Rlx}}\Z)(\Y-\widehat{\PhiB}^{\text{Rlx}}\Z)^{\top} 
\end{align}
Then, assuming $\widehat{\Sigma}_u$ is non-singular, the feasible GLS estimator can be expressed as
\begin{align}
\label{FGLS}
\vect(\widehat{\PhiB}^{FGLS})=\mathbf{R}[\mathbf{R}^{^{\top}}(\Z\Z^{\top}\otimes \widehat{\Sigma}_u^{-1})\mathbf{R}]^{-1}\mathbf{R}^{\top}(\Z\otimes \widehat{\Sigma}_u^{-1})\text{vec}(\Y).
\end{align}




Due to the poor numerical properties detailed by \cite{foschi1}, it is inadvisible to form \eqref{SigmaUEst} or \eqref{FGLS} directly.  In addition, our applications have found $\Z\Z^{\top}$ to be poorly conditioned when T is small.  
Moreover, as the dimension increases, conducting operations directly with the $(k^2p+k)\times (k^2p+k)$ matrix $(\Z\Z^{\top}\otimes I_k)$ exhausts memory.    

To ameliorate these issues of dimensionality, the refitting procedure can be conducted in parallel across rows of $\B$ if the covariance matrix is assumed to be the identity.  Additionally, the conditioning of $\Z\Z^{\top}$ can be improved by implementing a modification of the procedure developed by \cite{Neumaier} (discussed in Section \ref{secIC}), which adds a small (on the order of $\epsilon_{\text{machine}}$) regularization penalty to $\Z$ and $\Y$ and computes \eqref{RLS} via a QR factorization that does not require explicit matrix inversion (for details, see the Appendix of \cite{Nicholson}).  However, this approach cannot be extended to incorporate a non-identity covariance.  The following sections detail an iterative procedure that constructs the feasible GLS estimator \eqref{FGLS} without explicit matrix inversion.   

\subsection{Generalized Least Squares}
Consider the conventional least squares problem with design matrix X $\in \mathbb{R}^{m\times n}$ ($n<m$) of full rank, response vector $y\in \mathbb{R}^{m}$, and vector of unknown coefficients $\beta\in \mathbb{R}^{n}$.  
\begin{align*}
  y=X\beta+\epsilon\\
\epsilon\stackrel{\text{iid}}{\sim}N(0,\Sigma)
\end{align*}
Generalized Least Squares (see e.g. \cite{bjorck}) incorporates an $m\times m$ semidefinite symmetric matrix $\Sigma$, leading to the optimization problem
\begin{align*}
\min_{\beta} ||C^{-1}(X\beta-y)||_2, 
\end{align*}
in which $\Sigma=CC^{\top}$.
The normal equations for the GLS problem take the form
\begin{align}
\label{GLS1}
X^{\top}\Sigma^{-1}X\beta=X^{\top}\Sigma^{-1}y, 
\end{align}
Equation \eqref{GLS1} can be unstable if X or $\Sigma$ are ill-conditioned.  Instead of solving \eqref{GLS1} directly, \cite{foschithesis} 
recommend formulating the system as the generalized linear least squares problem (GLLSP)
\begin{align}
\label{GLLSP}
\argmin_{v,\beta} v^{\top}v  \text{ s.t. } y=X\beta+Cv, 
\end{align}
  in which $v\in \mathbb{R}^{m}, Cv=\epsilon$.  This problem can be solved with a generalized QR factorization \citep{Paige} which does not require explicit matrix inversion.  The generalized QR factorization involves the QR factorization of X 
\begin{align}
\label{QR1}
  &Q^{\top}X= \begin{blockarray}{c@{}c@{\hspace{4pt}}cl}
    \begin{block}{[c@{\hspace{5pt}}c@{\hspace{5pt}}c]l}
      & R  & & \mLabel{n} \\
      & 0  & & \mLabel{m-n}, \\
    \end{block}
  \end{blockarray}\\
\end{align}
in which $Q \in \mathbb{R}^{m\times m},$ is an orthogonal matrix and $R\in \mathbb{R}^{n\times n}$ is upper triangular, and the product RQ\footnote{ The RQ decomposition of X can be computed from the QR decomposition of $X^{\top}$ with the rows reversed.  The resulting R from the QR decomposition then needs to be transposed with its rows and columns reversed} decomposition of $Q^{\top}C$ 
  The product RQ decomposition takes the form 
\begin{align*}
&(Q^{\top}C)P =  U, \\
&U= \begin{blockarray}{c@{}cc@{\hspace{4pt}}cl}
    & n & m-n  & & \\
    \begin{block}{[c@{\hspace{5pt}}cc@{\hspace{5pt}}c]l}
      & U_{11} & U_{12}  & & \mLabel{n} \\
      & 0 & U_{22}  & & \mLabel{m-n} \\
    \end{block}
  \end{blockarray},  
\end{align*}
in which $P\in \mathbb{R}^{m\times m}$ is an orthogonal matrix and $U\in \mathbb{R}^{m\times m}$ is upper triangular.  Since P is orthogonal, $\|v\|_2=\|P^{\top}v\|_2$, so Equation \eqref{GLLSP} can be reformulated as:
\begin{align*}
&\min_{v,\beta}  \|P^{\top}v\|_2^2 \text{ s.t. } Q^{\top}y= Q^{\top}X\beta+Q^{\top}CPP^{\top}v,\\
\iff & \min_{v,\beta} \|P^{\top}v\|_2^{2} \text{ s.t. }  Q^{\top}y= Q^{\top}QR\beta+UP^{\top}v  
\end{align*}
next, define $P^{\top}v=(v_1^{\top},v_2^{\top})$, in which $v_1$ has length n and $v_2$ has length m-n.  Then, we can express the GLLSP as
\begin{align}
&\argmin_{v_1,v_2,\beta} \|v_1\|_2^2+\|v_2\|_2^2 \text{ subject to }\\
\label{y1eq}
&y_1=R\beta+U_{11}v_1+U_{12}v_2,\\
\label{y2eq}
&y_2=U_{22}v_2,  
\end{align}
in which 
\begin{align}
\label{QR2}
  &Q^{\top}y= \begin{blockarray}{c@{}c@{\hspace{4pt}}cl}
    \begin{block}{[c@{\hspace{5pt}}c@{\hspace{5pt}}c]l}
      & y_1  & & \mLabel{n} \\
      & y_2  & & \mLabel{m-n} \\
    \end{block}
  \end{blockarray}
\end{align}
 \cite{Paige} notes that since $R$ is of full rank, Equation \eqref{y1eq} can always be solved for $\beta$ once $v_1$ and $v_2$ are given, hence Equation \eqref{y2eq} gives the constraints on $v,$ reducing the problem to
\begin{align*}
&\argmin_{v} \|v_1\|_2^2+\|v_2\|_2^2 \\
&\text{ subject to }\\
&y_2=U_{22}v_2.  
\end{align*}
Now, since $v_1$ no longer appears in the constraints, we set it to zero and calculate $\hat{\beta}_{\text{FGLS}}$ by solving the triangular systems
\begin{align*}
R\beta&=y_1-U_{12}v_2,\\
U_{22}v_2&=y_2,  
\end{align*}
for $v_2$ and $\beta$.  We can then estimate $\widehat{\Sigma}$ as $v_2^{\top}v_2/(m-n)$.

\subsection{Application to Relaxed Feasible Generalized Least Squares}
\label{secifgls}
The previous approach can be modified to take into account the structure of the VAR.  \cite{Foschi3}, develop an extension of this procedure in the context of seemingly unrelated regressions with common regressors and later extend it to the VAR in \cite{foschi1}.  Their framework can easily be extended to our context of Relaxed VAR estimation.  We start by formulating the Generalized Linear Least Squares Problem 
\begin{align*}
&\argmin_{V} ||V||_F \\
&\text{ subject to } \vect(\Y^{\top})= ((I_k \otimes \Z^{\top})\mathbf{R})\phi_{\text{FGLS}} + \vect(V C^{\top}), 
\end{align*}
in which $\mathbf{R}$ is the $(k^2p+k)\times r$ restriction matrix of $\PhiB^{\top}$ and $\vect(\widehat{\PhiB}_{\text{FGLS}}^{\top})=\mathbf{R}\phi_{\text{FGLS}}$.  Note that this application uses $\Z^{\top}$ because its corresponding Kronecker product produces a block diagonal structure.   For notational ease, we will define $\mathbf{X}=(I_k \otimes \Z^{\top})\mathbf{R}$.  Here, $C$ is a lower triangular Cholesky factor such that $\Sigma_u=CC^{\top}$, and V is defined as a random matrix satisfying the relationship $(C\otimes I_T)\vect(V)=\vect(U)$.
\\
\\
First, note that we do not need to directly compute the QR factorization of $\X$.  We can instead compute the QR factorization of $\Z^{T}R_i$ for each $i=1,\dots,k$, in which $R_i$ is the $(kp+1)\times r_i$ restriction matrix for series i (denoted by a row in $\PhiB$).  The Q corresponding to $\mathbf{X}$ can then be formed as:
\begin{align}
\label{QRDECOMP}
\mathbf{Q} =\begin{pmatrix}
  \oplus_i \widetilde{Q}_i & \oplus_i \widehat{Q}_i   
\end{pmatrix}
=
\begin{pmatrix}
\widetilde{Q}_1 & & &\widehat{Q}_1 & & \\
&  \ddots & & &\ddots & \\
& &  \widetilde{Q}_k & & & \widehat{Q}_k,   
\end{pmatrix}
\end{align}
in which $\widetilde{Q}_i$ represents the ``economy'' $Q$ for series i (of dimension $T\times r_i$) and $\widehat{Q}_i$ represents the orthogonal completion for series $i$ (of dimension $T \times (kp+1-r_i)$).  The upper triangular matrix can be constructed similarly (to avoid confusion, we denote it as $\mathcal{R}$).  

Next, we construct
\begin{align*}
&\vect{(\widetilde{\Y})}=\widetilde{\mathbf{Q}}^{\top}\vect{(\Y)}\\
&\vect{(\widehat{\Y})}=\widehat{\mathbf{Q}}^{\top}\vect{(\Y)}.
\end{align*}
Now, consider the generalized QR decomposition:
\begin{align*}
&\mathbf{Q}^{\top}(C \otimes I_T)\mathbf{P}  =
\begin{blockarray}{cc@{}c@{\hspace{4pt}}cl}
    & r & kT-r  & & \\
    \begin{block}{[cc@{\hspace{5pt}}c@{\hspace{5pt}}c]l}
      & W_{11} & W_{12}   & & \mLabel{r} \\
      & 0 & W_{22}  & & \mLabel{kT-r} \\
    \end{block}
  \end{blockarray}  
\end{align*}
Using the above decomposition, the GLLSP becomes
\begin{align}
\label{RFGLS}
&\argmin_{\widetilde{v},\widehat{v}} ||\widetilde{v}||_2+||\widehat{v}||_2\\
&\text{ subject to }
\begin{pmatrix}
 \vect(\widetilde{\Y})\\
 \vect(\widehat{\Y})
\end{pmatrix}
=
\begin{pmatrix}
 \mathcal{R}\\
0 
\end{pmatrix}\phi_{\text{FGLS}} +
\begin{pmatrix}
 W_{11} & W_{12}\\
0 & W_{22} 
\end{pmatrix}
\begin{pmatrix}
 \vect(\widetilde{v})\\
 \vect(\widehat{v})  
\end{pmatrix}
\end{align}
in which
\begin{align*}
\mathbf{P}^{\top}\vect{(V)}=\begin{blockarray}{c@{}c@{\hspace{4pt}}cl}
    \begin{block}{[c@{\hspace{5pt}}c@{\hspace{5pt}}c]l}
      & \vect(\tilde{v})   & & \mLabel{r} \\
      & \vect(\hat{v})  & & \mLabel{kT-r} \\
    \end{block}
  \end{blockarray}.  
\end{align*}
Since, as in the univariate generalized least squares scenario, the constraints for $\tilde{v}$ are always satisfied, we set $\tilde{v}=0$ and solve the triangular system  
\begin{align*}
 W_{22}\vect{\hat{v}}=\vect{(\hat{y})}. 
\end{align*}
After doing so, we compute 
\begin{align*}
\vect{(v^{*})}=W_{12}\vect{(\hat{v})}.  
\end{align*}
Then, we solve the final triangular system
\begin{align*}
\mathcal{R}\phi_{\text{FGLS}}=\vect{(\widetilde{\Y})}-\vect{(v^{*})}  
\end{align*}

The RQ decomposition of $\mathbf{Q}^{\top}(C \otimes I_T )$ is the most computationally expensive component, requiring $O(k^3T^3)$ floating point operations.  \cite{Foschi3} conduct the RQ decomposition in two stages, first calculating
\begin{align}
\label{Kron}
 \mathbf{Q}^{\top}(C\otimes I_T)\mathbf{Q} =  \begin{blockarray}{cc@{}c@{\hspace{4pt}}cl}
    &r & kT-r  & & \\
    \begin{block}{[cc@{\hspace{5pt}}c@{\hspace{5pt}}c]l}
      & \widetilde{W}_{11} & \widetilde{W}_{12}   & & \mLabel{r} \\
      & \widetilde{W}_{21} & \widetilde{W}_{22}  & & \mLabel{kT-r} \\
    \end{block}
  \end{blockarray}.  
\end{align}
in which each $\widetilde{W}_{ij}$ is block upper triangular.  
In the second stage, the RQ decompositions is computed
\begin{align*}
 & \begin{pmatrix}
   \widetilde{W}_{21} & \widetilde{W}_{22}  
  \end{pmatrix}
\widetilde{P}=
\begin{pmatrix}
  0 & W_{22}
\end{pmatrix},
\end{align*}
and $W_{12}$ is constructed by using the previously calculated $\widetilde{P}$
\begin{align*}
 & \begin{pmatrix}
   \widetilde{W}_{11} & \widetilde{W}_{12}  
  \end{pmatrix}
\widetilde{\mathbf{P}}=
\begin{pmatrix}
  W_{11} & W_{12}
\end{pmatrix}.
\end{align*}
Therefore, we can conclude that $\mathbf{P}=\mathbf{Q}\widetilde{\mathbf{P}}$.  \cite{Foschi3} details efficient algorithms which take advantage of the Kronecker structure when computing these factorizations.  One can also avoid explicitly forming  $\mathbf{Q}^{\top}(C\otimes I_T)\mathbf{Q}$ by separately constructing each submatrix in Equation \ref{Kron} (derivations are in Section \ref{secadd}).  This procedure is summarized in Algorithm \ref{algIC} in Section \ref{tabalg}. 

\subsubsection{Updating $\widehat{\Sigma}_U$}
As $\Sigma_U$ is typically unknown, this approach is usually an iterative procedure.  Initially, $\widehat{\Sigma}_U$ is set to $I_k$, and is updated based on the model residuals $\widetilde{U}$, which are calculated as follows 
\begin{align*}
&\vect{(\widetilde{U})} = \widetilde{\Y}-(I_k \otimes R)\widehat{\phi}_{\text{FGLS}}\\
&\widehat{\Sigma}_{u}^{(j+1)}=\frac{\widetilde{U}^{\top}\widetilde{U}+ \widehat{\Y}^{\top}\widehat{\Y}}{T-(p\times k)} 
\end{align*}
One can update $\widehat{\Sigma}_u$ and $\widehat{\phi}_{\text{FGLS}}$ until some convergence criterion is satisfied; we currently use the matrix 2-norm (i.e. the maximum singular value of $\widehat{\Sigma}_u^{(j+1)}-\widehat{\Sigma}_u^{(j)}$).    Note that $\widehat{\Y}$ does not change across iterations, so the only component that needs to be updated is $\widetilde{U}$.

A potential complication arises when $\widehat{\Sigma}_u^{j+1}$ is not positive definite.  In this scenario, since it is not possible to take its Cholesky decomposition, the algorithm will break down.  \cite{Foschi3} propose computing the Cholesky factor $C$ directly from the QL decomposition of

\begin{align*}  
\begin{bmatrix}
\widetilde{U}\\
\widetilde{\Y}  
\end{bmatrix}.
\end{align*}

 However, this decomposition does not guarantee that the diagonal of $C$ will be positive.  As an alternative, in the rare instances when $\widehat{\Sigma}_u$ is not positive definite, we propose factoring $\widehat{\Sigma}_u$ using the singular value decomposition
 \begin{align*}
 &\widehat{\Sigma}_u=UDU^{\top}\\
 &Q,C=\text{QR}(D^{1/2}U^{\top}),  
 \end{align*}
i.e., $C$ is recovered from the $R$ in the QR decomposition of $D^{-1/2}U^{\top}$.
\subsection{Additional  Details}
\label{secadd}
\subsubsection{Construction of Submatrices in Equation \ref{Kron}}
Each of the 4 submatrices in Equation \ref{Kron} can be expressed in closed form; as combinations of the ``economy'' QR decompositions for each series $(\widetilde{Q}_i)$ as well as its orthogonal completion $(\widehat{Q}_i)$.  

\begin{align*}
W_{11}\in \mathbb{R}^{r\times r}=
&\begin{bmatrix}
C_{1,1}I_{r_1} &  C_{2,1}\widetilde{Q}_2^{\top}\widetilde{Q}_1 &\dots & \dots & C_{k,1}\widetilde{Q}_k^{\top}\widetilde{Q}_1\\
\utwi{0} & C_{2,2}I_{r_2} & C_{3,2}\widetilde{Q}_3^{\top}\widetilde{Q}_2 & \dots & C_{k,2}\widetilde{Q}_k^{\top}\widetilde{Q}_2\\
\vdots & \ddots & \ddots & \ddots & \ddots\\
\utwi{0} & \utwi{0} & \dots & C_{kk}I_{r_k} & \utwi{0}\\
\end{bmatrix}\\
W_{22}\in \mathbb{R}^{kT-r\times kT-r}=
&\begin{bmatrix}
C_{1,1}I_{T-r_1} &  C_{2,1}\widehat{Q}_2^{\top}\widehat{Q}_1 &\dots & \dots & C_{1,k}\widehat{Q}_k^{\top}\widehat{Q}_1\\
\utwi{0} & C_{2,2}I_{T-r_2} & C_{2,3}\widehat{Q}_3^{\top}\widehat{Q}_2 & \dots & C_{2,k}\widehat{Q}_k^{\top}\widehat{Q}_2\\
\vdots & \ddots & \ddots & \ddots & \ddots\\
\utwi{0} & \utwi{0} & \dots & C_{kk}I_{T-r_k} & \utwi{0}\\
\end{bmatrix}\\
W_{12}\in \mathbb{R}^{r \times kT-r}=
&\begin{bmatrix}
\utwi{0}_{r_1} &  C_{2,1}\widetilde{Q}_1^{\top}\widehat{Q}_2 &\dots & \dots & C_{1,k}\widetilde{Q}_1^{\top}\widehat{Q}_1\\
\utwi{0} & \utwi{0}_{r_2} & C_{2,3}\widetilde{Q}_2^{\top}\widehat{Q}_3 & \dots & C_{2,k}\widetilde{Q}_2^{\top}\widehat{Q}_k\\
\vdots & \ddots & \ddots & \ddots & \ddots\\
\utwi{0} & \utwi{0} & \dots & \dots & \utwi{0}\\
\end{bmatrix}\\
W_{21}\in \mathbb{R}^{kT-r\times r}=
&\begin{bmatrix}
\utwi{0}_{r_1} &  C_{2,1}\widetilde{Q}_2^{\top}\widehat{Q}_1 &\dots & \dots & C_{1,k}\widetilde{Q}_k^{\top}\widehat{Q}_1\\
\utwi{0} & \utwi{0}_{r_2} & C_{2,3}\widetilde{Q}_3^{\top}\widehat{Q}_2 & \dots & C_{2,k}\widetilde{Q}_k^{\top}\widehat{Q}_2\\
\vdots & \ddots & \ddots & \ddots & \ddots\\
\utwi{0} & \utwi{0} & \dots & \dots & \utwi{0}\\
\end{bmatrix}
\end{align*}
in which $r_i$ denotes the number of active coefficients for series i.

\begin{align*}
 \mathbf{Q}^{\top}(C\otimes I_T)\mathbf{Q} =  \begin{blockarray}{cc@{}c@{\hspace{4pt}}cl}
    &r & kT-r  & & \\
    \begin{block}{[cc@{\hspace{5pt}}c@{\hspace{5pt}}c]l}
      & \widetilde{W}_{11} & \widetilde{W}_{12}   & & \mLabel{r} \\
      & \widetilde{W}_{21} & \widetilde{W}_{22}  & & \mLabel{kT-r} \\
    \end{block}
  \end{blockarray}.  
\end{align*}



\subsection{Tables and Algorithms}
\label{tabalg}
     \begin{table}[H]
       \centering
           \caption{ \label{tab:tabStruct} Arguments for {\tt struct} in {\tt constructModel}.  X denotes ``True'' while . denotes ``False.'' }     
   \begin{tabular}{ |r | l| c | c | c}
           Struct Argument& Penalty & VAR Support & VARX Support & Univariate Support \\
    \hline
``Lag'' &  Lag Group & X & X  & X  \\
``OwnOther''&  Own/Other Group & X & X &. \\
``SparseLag'' & Lag Sparse Group & X & X &  . \\
``SparseOO''& O/O Sparse Group & X & X & . \\
``Basic &  Basic & X & X & X  \\  
``EF'' & Endogenous-First  & . & X  & .\\
``HVARC'' & Componentwise Hierarchical  & X & . & X \\ 
``HVAROO'' & Own/Other Hierarchical  & X & . & . \\ 
``HVARELEM'' & Elementwise Hierarchical  & X & . & . \\ 
``Tapered'' & Lag Weighted Lasso  & X & . & .\\ 
  \end{tabular}
 \end{table}

      \begin{table}[H]
    \centering   
           \caption{\label{tab:tabSA} Solution Algorithms employed for each structured penalty }     
   \begin{tabular}{ |l | l| l |}
Algorithm&  Solution Procedure & Reference \\
    \hline
 Lag & Block Coordinate Descent & \cite{Goldfarb}    \\
 Own/Other & Block Coordinate Descent& .  \\
Lag Sparse & Proximal Gradient Descent &  \cite{beck}   \\
 Own/Other Sparse & Proximal Gradient Descent& .  \\
 Basic & Coordinate Descent &  \cite{Friedman}  \\  
Endogenous-First & Fast Iterative Soft Thresholding& \cite{Jenatton}
 \end{tabular}
 \end{table}


\begin{algorithm}
\footnotesize
\caption{Iterative procedure to determine $\lambda_{\max}$}
\label{alg1}
\begin{algorithmic}[5]
\Require $\Y,\Z,\B,\lambda_{\max \text{coarse}},\epsilon$   
\State $\lambda_{\text{HIGH}}\leftarrow \lambda_{\max \text{coarse}}$
\State $\lambda_{\text{LOW}} \leftarrow 0$
\While{$\lambda_{\text{HIGH}}-\lambda_{\text{LOW}}>\epsilon$}
\State $\lambda \leftarrow \frac{\lambda_{\text{HIGH}}+\lambda_{\text{LOW}}}{2}$
\State $\B \leftarrow \text{BigVAR Model}(\Y,\Z,\B,\lambda)$
\If{$\|B\|_{\infty}=0$}
\State $\lambda_{\text{HIGH}}\leftarrow \lambda$
\Else
\State $\lambda_{\text{LOW}}\leftarrow \lambda$
\EndIf
\EndWhile
\Return $\lambda$
\end{algorithmic}
\end{algorithm}

\begin{algorithm}
\footnotesize
\caption{Fit a VARX according to information criterion minimization}
\label{algIC}
\begin{algorithmic}[5]
\Require $\Y,\Z,\B,p,s,\text{criterion}$   
\For{$i=0,\dots,p$}
\For{$j=0,\dots,s$}
\If{$i>0$ \& $j>0$}
\State $K=[\Z_{:,\big(1:(ki+1),(kp+2):((kp+2):js)\big)}^{\top},\Y^{\top}]$
\ElsIf{$i=0$ \& $j>0$ }
\State $K=[\Z_{:,(kp+2):(kp+2+js)}^{\top},\Y^{\top}]$
\ElsIf{$i>0$ \& $j=0$ }
\State $K=[\Z_{:,(1:(ki+1))}^{\top},\Y^{\top}]$
\ElsIf{$i=0$ \& $j=0$}
\State $K=[\utwi{1}]$
\EndIf
\State $\widehat{\Sigma}_u \leftarrow \text{VARXFit}(K)$
\If{criterion=``AIC''}
\State $IC[i,j]\leftarrow |\widehat{\Sigma}_u|+\frac{2(k(ki+mj+1))}{T-\max(i,j)}$
\ElsIf{criterion=``BIC''}
\State $IC[i,j]\leftarrow |\widehat{\Sigma}_u|+\frac{\log(T-\max(i,j))(k(ki+mj+1)}{T-\max(i,j)}$
\EndIf
\EndFor
\EndFor
\\
\Return $\hat{p},\hat{s}$ as the minimum entry of $IC[i,j]$
\Procedure{VARXFit}{K}
\State $Q,R \leftarrow QR(K)$
\State $R_{11}=R_{1:(kp+ms+1),1:(kp+ms+1)}$
\State $R_{12}=R_{1:(kp+ms+1),(kp+ms+1):(kp+ms+k+1)}$
\State $R_{22}=R_{(kp+ms+1):(kp+ms+k+1),(kp+ms+1):(kp+ms+k+1)}$
\State $\widehat{B}\leftarrow (R_{11}^{-1}R_{12})^{\top}$
\State $\widehat{\Sigma}_u\leftarrow \frac{R_{22}^{\top}R_{22}}{\text{nrow}(K)}$\\
\Return $\widehat{\B}$,$\widehat{\Sigma}_u$
\EndProcedure
\end{algorithmic}
\end{algorithm}

\begin{algorithm}
\footnotesize
\caption{Iterated Feasible GLS (adapted from Algorithm 21.1 of \cite{Foschi3})}
\label{algIC}
\begin{algorithmic}[5]
\Require $\Y,\X,\B,p,k,\Sigma_{u}^{(0)},\epsilon_1,\epsilon_2$   
\State $\mathbf{R}\leftarrow\text{RestrictionMatrix}(\B^{\top},\epsilon_1)$ \Comment{selects coefficients greater in magnitude than $\epsilon_1$.}
\State $\mathbf{Q}\leftarrow \text{determined according to Equation} \eqref{QRDECOMP}$
\State $\mathcal{R}\leftarrow \begin{pmatrix}
\mathcal{R}_1 & & \\
&  \ddots &  \\
& &  \mathcal{R}_k,   
\end{pmatrix}$

\State $\text{vec}(\widetilde{\Y})=\widetilde{\mathbf{Q}}^{\top}\Y$
\State $\text{vec}(\widehat{\Y})=\widehat{\mathbf{Q}}^{\top}\Y$
\State $C\leftarrow \text{Cholesky}(\Sigma_u^{(0)})$
\State $j\leftarrow 1$
\While{threshold$>\epsilon_2$}\\
\State $\mathbf{Q}^{\top}(C\otimes I_T)\mathbf{Q} =  \begin{blockarray}{cc@{}c@{\hspace{4pt}}cl}
    &r & kT-r  & & \\
    \begin{block}{[cc@{\hspace{5pt}}c@{\hspace{5pt}}c]l}
      & \widetilde{W}_{11} & \widetilde{W}_{12}   & & \mLabel{r} \\
      & \widetilde{W}_{21} & \widetilde{W}_{22}  & & \mLabel{kT-r} \\
    \end{block}
  \end{blockarray}.$
\State $
  \begin{pmatrix}
   \widetilde{W}_{21} & \widetilde{W}_{22}  
  \end{pmatrix}
\widetilde{\mathbf{P}}=
\begin{pmatrix}
  0 & W_{22}
\end{pmatrix}$
\State 
 $ \begin{pmatrix}
   \widetilde{W}_{11} & \widetilde{W}_{12}  
  \end{pmatrix}
\widetilde{\mathbf{P}}=
\begin{pmatrix}
  W_{11} & W_{12}
\end{pmatrix}.
$\Comment{RQ Decomposition}
\State Triangular Solve (for $v$) \;\;\; $W_{22}\text{vec}(v)=\text{vec}(\widehat{\Y})$
\State $v^{*}=W_{12}\text{vec}(v)$
\State Triangular Solve \;\;\; $\mathcal{R}\text{vec}(\widehat{\phi}^{IFGLS})=\text{vec}(\widetilde{\Y}-v)$
\State $\text{vec}(\widetilde{U})\leftarrow \widetilde{\Y}-(I_k\otimes \mathcal{R}_1)\text{vec}(\widehat{\phi}^{IFGLS})$
\State $\Sigma_u^{(j)}\leftarrow \frac{\widetilde{U}^{\top}\widetilde{U}+\widehat{\Y}^{\top}\widehat{\Y}}{T-(p\times k)}$
\State $\epsilon_2\leftarrow\|\Sigma_u^{(j)}-\Sigma_u^{(j-1)}\|_2$ \Comment{Operator Norm}
\State $C\leftarrow \text{Cholesky}(\Sigma_u^{(j)})$
\EndWhile
\\
\State $\widehat{\PhiB}^{(IFGLS)}\leftarrow \mathbf{R}\text{vec}(\widehat{\phi}^{IFGLS})$\\

\Return $\widehat{\PhiB}^{(IFGLS)},\Sigma_u^{(j)}$
\end{algorithmic}
\end{algorithm}

\end{document}